\pdfoutput=1

 \documentclass[12pt,preprint]{aastex}

\shorttitle{Activity Analyses for Solar-Type Stars}
\shortauthors{He et al.}

\begin{document}

\title{Activity Analyses for Solar-Type Stars Observed With Kepler. \\I. Proxies of Magnetic Activity}

\author{Han He, Huaning Wang, and Duo Yun}
\affil{Key Laboratory of Solar Activity, National Astronomical Observatories, Chinese Academy of Sciences, Beijing 100012, China.}
\email{hehan@nao.cas.cn}

\begin{abstract}
  Light curves of solar-type stars often show gradual fluctuations due to rotational modulation by magnetic features (starspots and faculae) on stellar surfaces. Two quantitative measures of modulated light curves are employed as the proxies of magnetic activity for solar-type stars observed with Kepler telescope. The first is named autocorrelation index $i_{\rm AC}$, which describes the degree of periodicity of the light curve; the second is the effective fluctuation range of the light curve $R_{\rm eff}$, which reflects the depth of rotational modulation. The two measures are complementary and depict different aspects of magnetic activities on solar-type stars. By using the two proxies $i_{\rm AC}$ and $R_{\rm eff}$, we analyzed activity properties of two carefully selected solar-type stars observed with Kepler (Kepler ID: 9766237 and 10864581), which have distinct rotational periods (14.7 vs. 6.0 days). We also applied the two measures to the Sun for a comparative study. The result shows that both the measures can reveal cyclic activity variations (referred to as $i_{\rm AC}$-cycle and $R_{\rm eff}$-cycle) on the two Kepler stars and the Sun. For the Kepler star with the faster rotation rate, $i_{\rm AC}$-cycle and $R_{\rm eff}$-cycle are in the same phase, while for the Sun (slower rotator), they are in the opposite phase. By comparing the solar light curve with simultaneous photospheric magnetograms, it is identified that the magnetic feature that causes the periodic light curve during solar minima is the faculae of the enhanced network region, which can also be a candidate of magnetic features that dominate the periodic light curves on the two Kepler stars.
\end{abstract}



\section{Introduction \label{sec-introduction}}

Kepler space telescope can obtain the light curves of the stars in the selected region of the sky with the wavelength band from 423 to 897 nm \citep{2010Sci...327..977B, 2010ApJ...713L..79K}. When investigating the solar-type stars that possess the convection zones and magnetic activities, two prominent phenomena can be expected in the observed light curves: one is the gradual fluctuation of the light curves as the result of inhomogeneous dark or bright features (e.g., dark starspots and bright faculae) crossing the observable disk of the star owing to stellar rotation, which is called the rotational modulation \citep{2011A&A...529A..89D}; another phenomenon is the stellar flare \citep{2000ApJ...529.1026S, 2010ARA&A..48..241B, 2013PASJ...65...49S}, which manifests as transient spikes standing on the gradual component of the light curves \citep{2011AJ....141...50W, 2012Natur.485..478M, 2013ApJS..209....5S}. The inhomogeneous features such as starspots and faculae on the surface of solar-type stars are caused by the stellar photospheric magnetic field \citep{2005LRSP....2....8B, 2009A&ARv..17..251S, 2012LRSP....9....1R}. Then the gradual fluctuation characteristics of the light curves can imply the global magnetic activity properties of the stars. The stellar flares, on the other hand, are eruptive phenomenon that happens in the stellar atmosphere around starspot regions as the result of sudden magnetic energy release through the magnetic reconnection process \citep{2010ARA&A..48..241B, 2011LRSP....8....6S}. In this paper, we focus on the approaches for analyzing the global magnetic activity properties of the solar-type stars observed with Kepler by examining the gradual fluctuation characteristics of the light curves. The flare activities observed with Kepler will be analyzed in a subsequent paper.

On the Sun, our nearest solar-type star, the dark sunspots correspond to the concentrated strong magnetic field that emerges on the photosphere, while the bright faculae correspond to the enhanced network magnetic field, which is dispersed over a much larger area \citep{2006RPPh...69..563S}. The sunspots appear in groups; each group of sunspots is called an active region, and the zone covered by faculae on the solar surface is called the enhanced network region \citep{2000A&A...353..380F}. Literature on modeling the solar irradiance variations \citep{1986ApJ...302..826F, 1998ApJ...492..390L, 2000A&A...353..380F} revealed that both the sunspots and faculae play the dominant roles in modulating the light curve of the Sun: the dark sunspots diminish the solar irradiance flux, while the bright faculae raise the flux. But the solar irradiance does not change regularly along with the solar rotation. Two factors complicate the shape of the light curve. The first factor is the time evolution of the magnetic features. It had been found that the lifetimes of most active regions (sunspot groups) are shorter than the rotation period of the Sun \citep{1997SoPh..176..249P}, while the enhanced network regions (facula zones) favour a relatively longer lifetime for its larger spatial scale \citep{1985AuJPh..38..961Z, 2000Natur.408..445S}. The successive appearances and disappearances of magnetic features with different spatial scales and different lifetimes on the surface of the Sun disturb the rotational modulation and cause the irregular fluctuations of light curves \citep{2003A&A...403.1135L}. On the other hand, the global coverage of both sunspots and facula zones (and hence the solar irradiance, whose long length-scale trend is dominated by faculae) shows long-term periodical variation along with the 11 yr solar cycle \citep{2004A&ARv..12..273F}. The second factor that affects the appearance of the light curve is the differential rotation \citep{1984ARA&A..22..131H}. That is, the magnetic features on the surface of the Sun (both sunspots and faculae) rotate faster at low latitude and slower at high latitude. Thus, magnetic features at different latitudes may cause rotational modulations to the light curve with different fine timescales.

Those complexities related to the light-curve modulations by magnetic features on the Sun can also exist on other solar-type stars, though maybe with different manifestations. For example, the dominant magnetic feature that affects stellar radiant flux may change between starspots and faculae for different solar-type stars or for different epochs of stars \citep[e.g.,][]{1990ApJ...348..703D, 1998ApJS..118..239R, 2007ApJS..171..260L}; it is also possible that there exist long-lifetime starspots with much larger spatial scale (than sunspots) on certain solar-type stars \citep[e.g.,][]{1998A&A...330..685S, 2009A&ARv..17..251S}, which may cause very different light curves from those of the Sun. To characterize the activity properties of the magnetic features on the solar-type stars observed with Kepler, one has to employ certain quantitative measures of the Kepler light curves as the proxies of magnetic activity.

One commonly used measure of light curves as the proxy of stellar magnetic activity is the range or amplitude of light-curve fluctuation, which quantifies the depth of rotational modulation. It is believed that the deeper rotational modulation and larger amplitude of a light curve usually imply the bigger magnetic features (i.e., darker starspot area or brighter facula zone) on the star's surface \citep{2010ApJ...713L.155B}. This measure of stellar magnetic activity via the fluctuation range of light curves has been extensively employed in the literature for analyzing the Kepler data, but with different definitions. For example, in the articles by \citet{2011AJ....141...20B, 2013ApJ...769...37B}, all the intensity points of a given light curve are sorted and the difference between the $5\%$ and $95\%$ values (for avoiding anomalous data points) is taken as the measurement of the fluctuation range, whereas \citet{2010Sci...329.1032G} and \citet{2011ApJ...732L...5C} calculated the standard deviation (i.e., rms) of the light curve to indicate the amplitude of light-curve fluctuation.

In this work, we make further effort to study the light-curve characteristics associated with the magnetic activities on the solar-type stars observed with Kepler. Specifically, we explore the proxy of magnetic activity from another point of view, that is, the light curve's periodicity caused by the rotational modulation.

In fact, by measuring the fluctuation periods of the observed light curves (through Fourier or spectrum analysis techniques), rotational periods of solar-type stars can be determined with good precision and accuracy \citep{2010ARA&A..48..581S}. Since the magnetic features that modulate the light curves may be located at different latitudes, it is also possible to derive the differential rotation information from the minor differences between the period values detected for the same star \citep{2013A&A...560A...4R, 2014A&A...564A..50L}. But because of the time evolution of magnetic features (see description above for the Sun), the rotational modulation can be disturbed, and not all stars show periodic light curves at a given time \citep{2010ARA&A..48..581S}. In the worst cases, the rotation periods of the stars can be hardly determined from the light curves by the usual Fourier or spectrum analysis techniques. This situation is not rare for the solar-type stars observed with Kepler \citep{2013A&A...560A...4R}. In the general circumstances, the periodic property of a star's light curve may vary with time: when the configuration of the magnetic features on the surface of the star changes faster (relative to the rotation period of the star), the rotational modulation is severely disturbed, and then the light curve fluctuates irregularly and thus shows weaker periodicity; when the configuration of the magnetic features changes slower, the light curve fluctuates more regularly and thus exhibits relatively stronger periodicity.

Since the degree of periodicity of a light curve can reflect the time evolution information of the magnetic features, we propose it as a new proxy of magnetic activity for solar-type stars. In this paper, we mainly utilize this measure and the fluctuation range of light curves (as mentioned above) to analyze the activity variations of Kepler stars. Both measures originate from the phenomenon of rotational modulation in light curves, yet they manifest different aspects of magnetic activities: the periodicity of a light curve describes the regularity of the rotational modulation and is related to the stability (or, conversely, the rate of change) of magnetic features, whereas the fluctuation range of light curves indicates the depth of the rotational modulation and is related to the magnetic features' size and spatial distribution. The two proxies are complementary and just like two orthogonal coordinates that can be used to depict the stellar magnetic activity characteristics concealed in the light curves.

We employ the autocorrelation algorithm \citep[e.g.,][]{2002.book.Brockwell} and introduce an autocorrelation index, $i_{\rm AC}$, to quantitatively describe the degree of periodicity of light curves. The autocorrelation-based method has proved to be a reasonable approach for determining the rotational periods of Kepler stars \citep{2013MNRAS.432.1203M, 2014ApJS..211...24M, 2014A&A...564A..50L}. We explain the concept and algorithm of $i_{\rm AC}$ in Section \ref{sec-Proxies}. For the second proxy of magnetic activity, we define a quantity, $R_{\rm eff}$, to represent the effective range of light-curve fluctuation. $R_{\rm eff}$ is based on the rms algorithm as in the work by \citet{2010Sci...329.1032G} and \citet{2011ApJ...732L...5C}, but with a slightly different definition. The evaluation of $R_{\rm eff}$ is also described in Section \ref{sec-Proxies}.

As an application of the two proxies of magnetic activity, by using the two quantitative measures $i_{\rm AC}$ and $R_{\rm eff}$, we investigate activity variations of two carefully selected solar-type stars observed with Kepler. The two solar-type stars (Kepler ID: 9766237 and 10864581) have distinct rotational periods: the first one is about 14.7 days (Earth day; the same below), and the second one is about 6.0 days. It is expected that stars with different rotation periods may manifest different activity behaviors \citep{1984ApJ...279..763N, 2007LRSP....4....3G, 2010ARA&A..48..581S}. The activity analyses for the two solar-type stars observed with Kepler (as well as the process of Kepler star selection and the procedure of data reduction) are described in detail in Section \ref{sec-activity-stars}. We also applied the two proxies of magnetic activity to the Sun for a comparative study. The light-curve data of the Sun employed in this research were obtained by the Solar and Heliospheric Observatory \citep[SOHO;][]{1995SoPh..162....1D}. The activity analysis for the Sun is given in Section \ref{sec-activity-sun}. In Section \ref{sec-discussion}, we discuss further the results of the above analyses. Section \ref{sec-summary-conclusion} provides a summary and conclusion.

\section{Proxies of Magnetic Activity \label{sec-Proxies}}

\subsection{Measurement of the Periodicity of the Light Curve \label{subsec-iAC}}

If a light curve shows some extent of periodicity, it means that the light curve has self-similarity after certain time lags. This self-similarity can be quantitatively measured by the correlation coefficient between the two copies of the time series data with and without the time lag. A sequence of correlation coefficients corresponding to a sequence of time lags makes up the autocorrelation function (ACF) of the light curve.

For an observed light-curve time series $\{X_t, t=0, 1, \ldots, N-1 \}$, where $N$ is the total number of the data points, the autocorrelation coefficient $\rho$ at time lag $h$ (i.e., the ACF) is defined by \citep{2002.book.Brockwell}
\begin{equation} \label{equ:rho-h}
\rho(h)=\frac{\sum_{t=0}^{N-1-h} (X_{t+h}-\bar{X})(X_t-\bar{X})}{\sum_{t=0}^{N-1} (X_t-\bar{X})^2},
\qquad 0\leqslant h \leqslant N-1,
\end{equation}
where $\bar{X}$ is the mean value of $\{X_t\}$:
\begin{equation}
\bar{X}=\frac{1}{N} \sum_{t=0}^{N-1} X_t.
\end{equation}

\begin{figure}
  \epsscale{0.96}
  \plotone{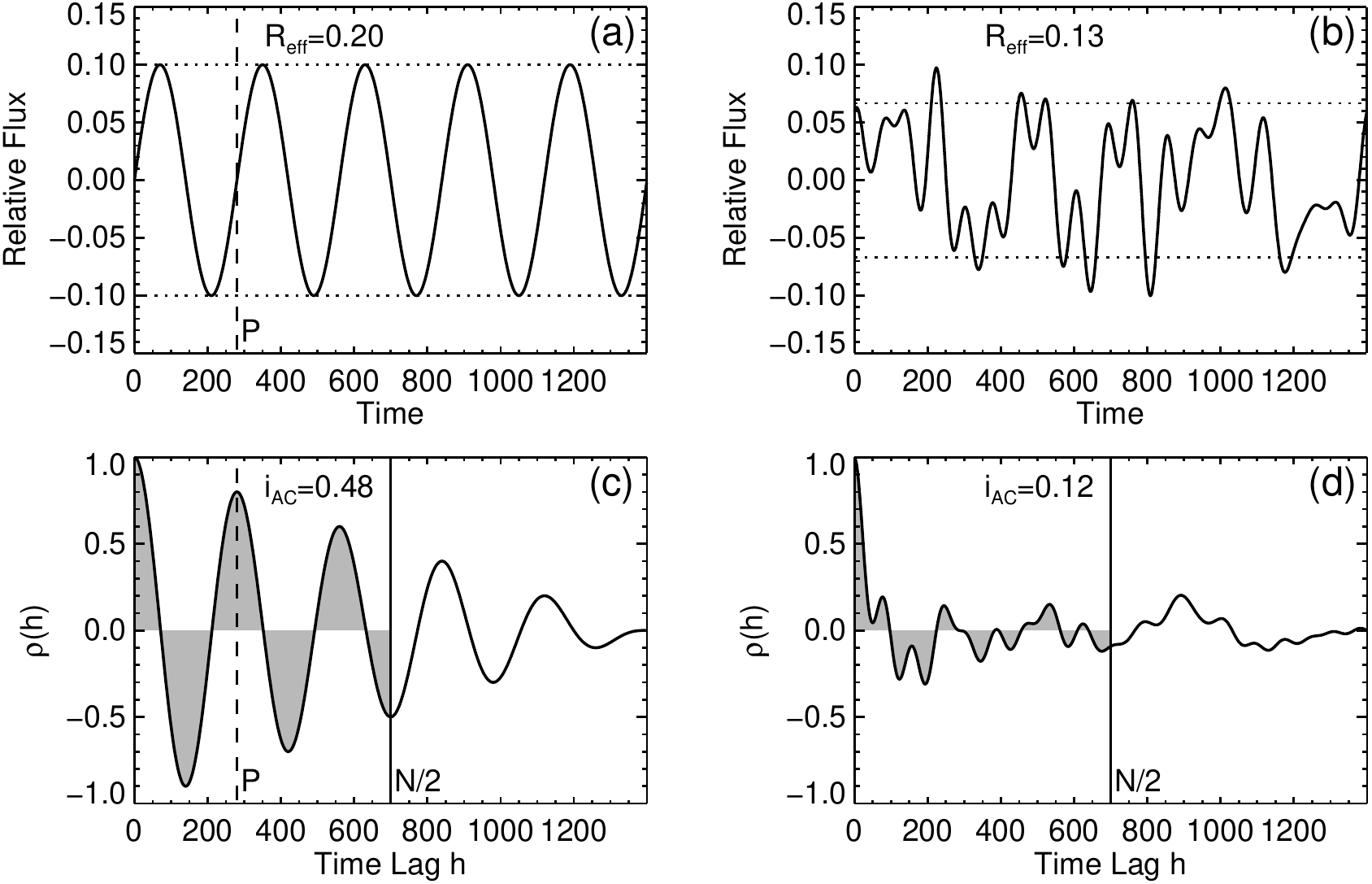}
  \caption{Diagram illustrating the concepts of $i_{\rm AC}$ and $R_{\rm eff}$ through two simulated light curves. (a) Regular sine curve with high periodicity. (b) Random fluctuation curve with poor periodicity. (c) Plot of the ACF, $\rho(h)$ (see equation (\ref{equ:rho-h})), for the sine curve. (d) Plot of $\rho(h)$ for the random curve. The two light curves in Figures \ref{fig01}a and \ref{fig01}b are in relative flux expression (see equation (\ref{equ:x_t})). The vertical dashed lines in Figures \ref{fig01}a and \ref{fig01}c indicate the period of the sine curve. The values of the autocorrelation index $i_{\rm AC}$ (see definition equation (\ref{equ:iAC})) for the two demo curves are given in Figures \ref{fig01}c and \ref{fig01}d, respectively; the shaded areas in the two $\rho(h)$ plots illustrate the integral term ($I_{\rm AC}$) in equation (\ref{equ:iAC}), which is proportional to the final value of $i_{\rm AC}$. The values of the effective fluctuation range $R_{\rm eff}$ (see defination equation (\ref{equ:R_eff})) for the two demo curves are given in Figures \ref{fig01}a and \ref{fig01}b, respectively; the two horizontal dotted lines in each light-curve plot illustrate the positions of $R_{\rm eff}/2$ and $-R_{\rm eff}/2$, and the value of $R_{\rm eff}$ can be understood geometrically as the distance between the two dotted lines.
  \label{fig01}}
\end{figure}

To demonstrate the properties of ACF, in Figure \ref{fig01} we show the $\rho(h)$ plots for two simulated light curves. One is a regular sine curve with high periodicity, and another is a random curve with poor periodicity. It can be seen in Figure \ref{fig01} that the value of $\rho$ is normalized into the interval of $(-1, 1]$ and the profiles of $\rho(h)$ fluctuate with the increase of time lag $h$. For the light curve with high periodicity (the sine curve in Figure \ref{fig01}a), the fluctuation amplitude of the ACF is large (see Figure \ref{fig01}c), implying strong autocorrelation. For the light curve with poor periodicity (see Figure \ref{fig01}b), the fluctuation amplitude of the ACF is relativity small (see Figure \ref{fig01}d), implying weak autocorrelation. Moreover, for the light curve with high periodicity, the $\rho(h)$ profile also shows regularly periodic variation, and the period of the $\rho(h)$ profile is the same as the period of the original light curve (illustrated by vertical dashed lines in Figures \ref{fig01}a and \ref{fig01}c) \citep{2013MNRAS.432.1203M}; for the light curve with poor periodicity, the $\rho(h)$ profile shows irregular variation (see Figures \ref{fig01}b and \ref{fig01}d).

To measure the degree of periodicity of a light curve with one quantity, we introduce the concept of an autocorrelation index (represented by $i_{\rm AC}$), which is defined as the average value of $|\rho(h)|$ over the interval $0 < h \leqslant N/2$:
\begin{equation} \label{equ:iAC}
i_{\rm AC}=\frac{2}{N} \int_{0}^{N/2} |\rho(h)| dh = \frac{2}{N} I_{\rm AC}.
\end{equation}
(If $N/2$ is not a natural number, $\rho(N/2)$ is computed via interpolation.) We choose $h \leqslant N/2$ because if $h > N/2$, only part of the light curve's data points are employed to calculate $\rho(h)$ (see equation (\ref{equ:rho-h})), which cannot reflect the global property of the light curve. The evaluated values of $i_{\rm AC}$ for the two demo light curves in Figure \ref{fig01} are given in Figures \ref{fig01}c and \ref{fig01}d, respectively. The integral term in equation (\ref{equ:iAC}) (represented by $I_{\rm AC}$) can be understood geometrically as the total area of the region enclosed by the ACF curve $\rho=\rho(h)$, the horizontal line $\rho=0$, and the vertical line $h=N/2$, as illustrated by the shaded areas in Figures \ref{fig01}c and \ref{fig01}d.

It can be seen in Figure \ref{fig01} that the light curve with stronger periodicity has a larger $i_{\rm AC}$ value (left column in Figure \ref{fig01}) and the light curve with weaker periodicity has a smaller $i_{\rm AC}$ value (right column in Figure \ref{fig01}). Thus, the autocorrelation index $i_{\rm AC}$ defined in equation (\ref{equ:iAC}) can be used as a quantitative measure to evaluate and compare the periodicity properties of different light curves. For a pure sine curve $x=A\sin(\omega t)$ as illustrated in Figure \ref{fig01}a, the corresponding ACF can be approximately expressed by equation $\rho(h)=(1-h/T)\cos(\omega h)$, where $T$ is the total length of the curve, and the value of $i_{\rm AC}$ approaches $3/(2\pi)\approx0.48$ when $T$ covers sufficient periods (see Figure \ref{fig01}c). For general curves with distortions and irregular fluctuations, the values of $i_{\rm AC}$ are expected to be in the interval between 0 and 0.48.

\subsection{Measurement of the Fluctuation Range of the Light Curve \label{subsec-Reff}}

Since fluctuation amplitude of a stellar light curve is generally not uniform, we define a quantity, $R_{\rm eff}$, to represent the effective range of the light-curve fluctuation. (In this work, when we refer to the fluctuation range of a light curve, we mean the distance between the crest and trough of the light-curve profile as adopted in the papers by \citet{2011AJ....141...20B, 2013ApJ...769...37B}.)

For a given light-curve time series $\{X_t, t=0, 1, \ldots, N-1 \}$, we first obtain the relative flux expression $\{x_t, t=0, 1, \ldots, N-1 \}$ of the light curve:
\begin{equation}\label{equ:x_t}
x_t=\frac{X_t-\widetilde{X}}{\widetilde{X}}, \qquad \widetilde{X}=\mathrm{median}(\{X_t\}).
\end{equation}
Equation (\ref{equ:x_t}) means that the original light-curve data $\{X_t\}$ are subtracted and then divided by the median value of $\{X_t\}$ (denoted by $\widetilde{X}$). The resulting relative flux data, $\{x_t\}$, fluctuate around the zero value, which is convenient for the further analysis. In fact, the light-curve plots shown in Figures \ref{fig01}a and \ref{fig01}b are based on the relative flux data. (Note that for a given light curve, the $\{X_t\}$ data and the corresponding $\{x_t\}$ data yield the same values of $\rho(h)$ and $i_{\rm AC}$ according to equations (\ref{equ:rho-h}) and (\ref{equ:iAC}).)

The effective fluctuation range of the light curve, $R_{\rm eff}$, is defined based on the relative flux data $\{x_t\}$:
\begin{equation}\label{equ:R_eff}
R_{\rm eff}=2(\sqrt{2} \cdot x_{\rm rms}),
\end{equation}
where $x_{\rm rms}$ is the rms value of $\{x_t\}$ \citep{2010Sci...329.1032G, 2011ApJ...732L...5C},
\begin{equation}\label{equ:x_rms}
x_{\rm rms}=\sqrt{\frac{1}{N} \sum_{t=0}^{N-1} x_t^2}.
\end{equation}

The evaluated $R_{\rm eff}$ values for the two demo light curves in Figure \ref{fig01} are given in Figures \ref{fig01}a and \ref{fig01}b, respectively; for each light curve, we use two horizontal dotted lines (upper and lower lines) to illustrate the positions of $R_{\rm eff}/2$ and $-R_{\rm eff}/2$, and the value of $R_{\rm eff}$ can be understood geometrically as the distance between the two dotted lines.

For the pure sine curve $x=A\sin(\omega t)$ as displayed in Figure \ref{fig01}a, $x_{\rm rms}=A/\sqrt{2}$, and $R_{\rm eff}=2 \cdot (\sqrt{2} \cdot A/\sqrt{2})=2A$ which is just the distance between the crest and trough of the light-curve profile (indicated by the two horizontal dotted lines in Figure \ref{fig01}a). That is why we introduce the factor $2\sqrt{2}$ in equation (\ref{equ:R_eff}). For general light curves with variable amplitudes, $R_{\rm eff}$ gives a quantitative assessment of the effective (i.e., balanced) range of fluctuation as illustrated in Figure \ref{fig01}b.

\section{Activity Analyses for Two Solar-Type Stars Observed with Kepler \label{sec-activity-stars}}

\subsection{Kepler Star Selection \label{subsec-star-selection}}

Two solar-type stars (G-type main-sequence stars) observed with Kepler were carefully selected for activity analysis by using the two proxies of magnetic activity. We selected the star samples based on the Kepler superflare star list given by \citet{2013ApJS..209....5S}; the stars in this list (279 G-type stars in total) had been identified with superflare spikes (flare energy greater than $10^{33}$ erg) in their light curves \citep{2013ApJ...771..127N, 2013ApJS..209....5S, 2014PASJ...66L...4N}, which ensures that the magnetic activities do exist on these stars. The criteria for our Kepler star selection from the list of \citet{2013ApJS..209....5S} are as follows:

\begin{enumerate}
  \item The dynamic range (upper limit value of $R_{\rm eff}$) of the star's light curve (in relative flux expression) should have the same order of magnitude as the dynamic range of the Sun (about $10^{-3}$; see Section \ref{sec-activity-sun}), since we want a comparative study between the Sun and the solar-type star samples.
  \item The light curve of the star should be continuously observed and have multiple epochs of regular fluctuations with stronger periodicity, as well as epochs of irregular fluctuations with weaker periodicity; then it will be favorable for analysis by using the $i_{\rm AC}$ measure.
  \item The selected stars should have distinctly different rotation periods, for we expect that the stars with different rotational periods may manifest different magnetic activity behaviors \citep{1984ApJ...279..763N, 2007LRSP....4....3G, 2010ARA&A..48..581S}.
\end{enumerate}

We surveyed all the light curves of the candidate Kepler stars in the list of \citet{2013ApJS..209....5S} and finally selected the two representative solar-type stars (Kepler ID: 9766237 and 10864581) that satisfy all the criteria given above. As required by the third criterion, the two selected stars have distinct rotation periods (denoted by $P_{\rm rot}$); the first is about 14.7 days, and the second is about 6.0 days (see Section \ref{subsec-stars-rotation-period} for details of the process to derive the $P_{\rm rot}$ values of the two stars).

For ease of reference, the first selected star (Kepler ID: 9766237) is hereafter referred to as Star I and the second star (Kepler ID: 10864581) is referred to as Star II. Table \ref{tb-basic-info-star12} gives the basic information (effective temperature, $T_{\rm eff}$; surface gravity, $\log g$; metallicity, [Fe/H]; radius, $R/R_\sun$; Kepler magnitude, $K_p$; and rotation period, $P_{\rm rot}$) of the two selected solar-type stars. The values of the stellar parameters ($T_{\rm eff}$, $\log g$, [Fe/H], $R/R_\sun$, $K_p$) in Table \ref{tb-basic-info-star12} are taken from the Kepler Input Catalog \citep[KIC;][]{2011AJ....142..112B}.

\subsection{Kepler Data Reduction \label{subsec-data-reduction}}
The light-curve data of the two selected solar-type stars were obtained by the photometry instrument aboard the Kepler spacecraft \citep{2010ApJ...713L..79K} and were acquired in long-cadence (LC) mode of the instrument \citep{2010ApJ...713L.120J}, i.e., one data point every 29.4 minutes. The raw flux data were processed by the Presearch Data Conditioning (PDC) module \citep{2012PASP..124.1000S, 2012PASP..124..985S} of the Kepler data analysis pipeline \citep{2010ApJ...713L..87J} to correct the systematic errors while keeping the astrophysical signals. The flux data after PDC processing (PDC flux, for short) are ready for the physical analyses. The latest PDC flux data product, Kepler Data Release 24 \citep{2015.KDR24Notes}, is utilized in this paper.

\begin{figure}
  \epsscale{1.0}
  \plotone{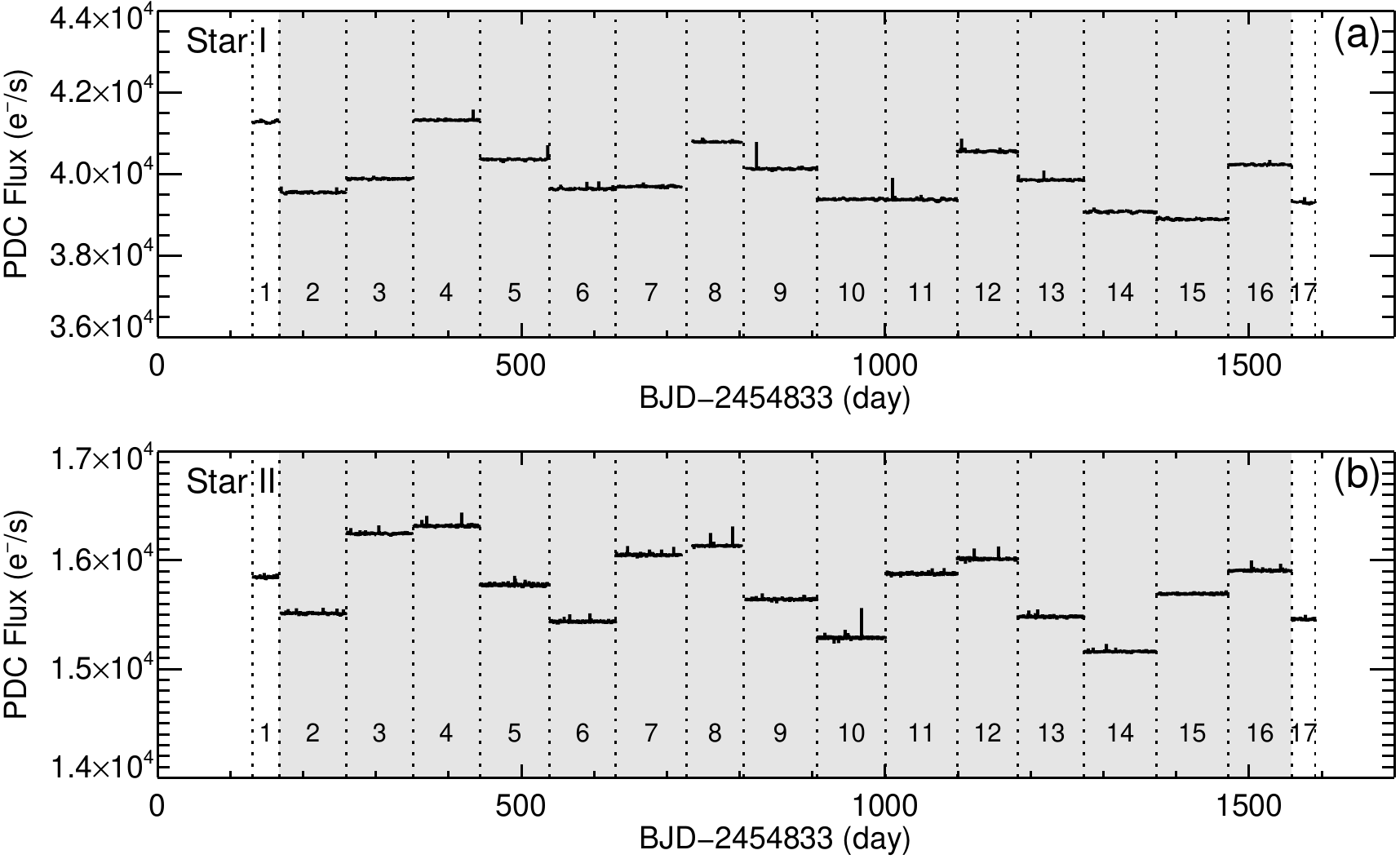}
  \caption{Overview of all the Kepler PDC flux data available for Star I (upper panel) and Star II (lower panel). These data are taken from the Kepler Data Release 24 \citep{2015.KDR24Notes}. The PDC flux curves of different quarters are separated by vertical dotted lines, and the quarter numbers are marked below the corresponding flux curves. The quarters highlighted in gray (Q2--Q16) are the full-length quarters that are utilized in this work for the magnetic activity analyses. BJD of the time axis refers to Barycentric Julian Date, and the offset 2,454,833 is the Julian Date at midday on 2009 January 1 \citep{2013.KDCHandbook}.
  \label{fig02}}
\end{figure}

The raw LC data and the processed PDC flux data of Kepler are organized by the observing quarters of the Kepler mission. A full-length quarter is about 3 months \citep{2010ApJ...713L.115H}. Figure \ref{fig02} gives an overview of all the PDC flux data available for Star I and Star II. The PDC flux curves of different quarters are separated by vertical dotted lines in Figure \ref{fig02} and the quarter numbers are marked below the corresponding flux curves. As shown in Figure \ref{fig02}, the flux data of both stars cover quarters 1--17 (Q1--Q17), in which Q2--Q16 are full-length quarters. In this work, we use the data of Q2--Q16 (15 quarters in total, highlighted in gray in Figure \ref{fig02}) for magnetic activity analyses, since Q1 and Q17 are too short to give a compatible result with other quarters (see also the unfolded light curves in Appendix \ref{sec-appendix-A}). The time system for the Kepler flux data is BJD$-$2,454,833 (which is adopted by the Kepler team; see the time axis in Figure \ref{fig02}), where BJD refers to the Barycentric Julian Date and the offset 2,454,833 is the value of the Julian Date at midday on 2009 January 1 \citep{2013.KDCHandbook}.

Because the Kepler telescope rolls $90^\circ$ about its axis between quarters and the photometry data of successive quarters for the same star are obtained by the different CCD modules \citep{2010ApJ...713L.115H}, the PDC flux plots in Figure \ref{fig02} show obvious discontinuities (shifts of the absolute values) between neighboring quarters and the long length-scale trends of the flux variations are not preserved in the PDC data \citep{2013.KDCHandbook}. For this reason, we processed the PDC flux data of different quarters separately. The reduction procedure for the PDC flux data in each quarter includes the following steps.

First, we obtained the relative flux expression (denoted by $f$) of the light curve from the original PDC flux data (denoted by $F$) based on equation (\ref{equ:x_t}),
\begin{equation}\label{equ:f}
f=\frac{F-\widetilde{F}}{\widetilde{F}}, \qquad \widetilde{F}=\mathrm{median}(F).
\end{equation}
The relative flux $f$ has the advantages that it fluctuates around zero and is not affected by the absolute value of the measured flux $F$. Thus, it can dispel the shift issues in the PDC flux data of different quarters and is also suitable for the comparative analysis between solar-type stars with different apparent magnitudes. (The PDC flux data, on the other hand, do rely on the stars' apparent magnitudes; see different $K_p$ values of Star I and Star II in Table \ref{tb-basic-info-star12} and the different PDC flux data ranges of the two stars in Figure \ref{fig02}.)

Second, the relative flux data $f$ were processed by a Fourier-based low-pass filter to remove the transient variation component (denoted by $f_{\rm T}$) of $f$, and we obtained the pure gradual variation component (denoted by $f_{\rm G}$) of the light curve. The transient variation component $f_{\rm T}$, which consists of noises, outliers, flare spikes, and granulation-driven flickers \citep[e.g.,][]{2014ApJ...781..124C, 2014A&A...570A..41K}, can disturb the follow-up activity analysis based on the light-curve measures $i_{\rm AC}$ and $R_{\rm eff}$ and thus is filtered out:
\begin{equation}
f_{\rm G}=f-f_{\rm T}.
\end{equation}
Since the observed flux data of different stars may have different noise levels and fluctuation properties, the upper cutoff frequencies of the low-pass filter for deriving $f_{\rm G}$ were determined via visual inspection. In this work, the values of the upper cutoff frequencies adopted for Star I and Star II approximate to $1/(0.3P_{\rm rot})$.

Third, based on the $f_{\rm G}$ time series data, the ACF of the light curve, $\rho(h)$, was derived by using equation (\ref{equ:rho-h}) and then the autocorrelation index, $i_{\rm AC}$, was evaluated by using equation (\ref{equ:iAC}) (see Section \ref{subsec-iAC} for details); meanwhile, the effective fluctuation range of the light curve, $R_{\rm eff}$, was acquired by substituting $f_{\rm G}$ into equation (\ref{equ:R_eff}) (see Section \ref{subsec-Reff} for details). Finally, we obtained the values of the two quantitative measures of the light curve, $i_{\rm AC}$ and $R_{\rm eff}$, which serve as the proxies of magnetic activity of the host star.

Figure \ref{fig03} illustrates the data reduction procedure for one quarter of PDC flux data using Q7 of Star I as the example. The time unit used in Figure \ref{fig03} is long-cadence \citep[i.e., sequence number of the data point, 1 long-cadence $\approx$ 29.4 minutes;][]{2010ApJ...713L.120J}, and the time axis is counted from the beginning of the quarter. Figure \ref{fig03}a displays the original PDC flux curve of the quarter, and Figure \ref{fig03}b displays the curves of the relative flux $f$ (gray color) and the gradual variation component of the relative flux $f_{\rm G}$ (black color). It can be seen in Figures \ref{fig03}a and \ref{fig03}b that the noises and outliers are evident in the original PDC flux data and the relative flux data $f$. These transient variations ($f_{\rm T}$ component) are well removed in the $f_{\rm G}$ flux curve as demonstrated in Figure \ref{fig03}b. Figure \ref{fig03}c displays the ACF of the light curve, $\rho(h)$, derived from the $f_{\rm G}$ time series data, and the evaluated value of $i_{\rm AC}$ is shown above the $\rho(h)$ curve. The evaluated value of $R_{\rm eff}$ based on the $f_{\rm G}$ data is given in Figure \ref{fig03}b, which is also illustrated by the two horizontal dotted lines (indicating the positions of $R_{\rm eff}/2$ and $-R_{\rm eff}/2$, respectively) in Figure \ref{fig03}b.

After all the data reduction processes described above, for each full-length quarter of light-curve data of Star I and Star II, we have the original PDC flux $F$, the relative flux $f$, and the gradual variation component of the relative flux $f_{\rm G}$, as well as one autocorrelation index value of the light curve, $i_{\rm AC}$, and one effective fluctuation range value of the light curve, $R_{\rm eff}$. The activity property analyses for the two selected solar-type stars in the next subsections are based on these data.

\subsection{Activity Properties of the Two Selected Solar-type Stars \label{subsec-stars-activity-properties}}

As explained in Section \ref{subsec-data-reduction}, we use the Kepler data of Q2--Q16 (i.e., full-length quarters) for analyzing the activity properties of the two selected solar-type stars. Figure \ref{fig04} gives an overview of the whole curves of the relative flux $f$ and the gradual variation component of the relative flux $f_{\rm G}$ within Q2--Q16 for Star I, which were obtained during the data reduction process described in Section \ref{subsec-data-reduction}. Figure \ref{fig05} displays the whole $f$ and $f_{\rm G}$ curves for Star II. In Figures \ref{fig04} and \ref{fig05}, the flux curves of different quarters are separated by vertical dotted lines, and the quarter numbers are marked below the corresponding curves.

It can be seen in Figures \ref{fig04} and \ref{fig05} that, by introducing the relative flux $f$, the light curves of different quarters are aligned to the same baseline (see also the unfolded curves of $f$ for Star I and Star II in Appendix \ref{sec-appendix-A}), which is more suitable for comparison between quarters than the original PDC flux shown in Figure \ref{fig02}. It can also be seen in Figures \ref{fig04} and \ref{fig05} that, by filtering out the transient variation component (i.e., noises, outliers, etc.) from the relative flux $f$ (see data reduction process in Section \ref{subsec-data-reduction}), the resulting $f_{\rm G}$ flux data can reflect the gradual fluctuation characteristics of the stellar light curves more clearly. As explained in Section \ref{sec-introduction}, it is the gradual variation component of a stellar light curve that is related to the rotational modulation and thus encompasses the activity information of the magnetic features on the star's surface. For this reason, we evaluated the two quantitative measures of light curves (proxies of magnetic activity), $i_{\rm AC}$ and $R_{\rm eff}$, based on the $f_{\rm G}$ data instead of $f$ as described in Section \ref{subsec-data-reduction}.

The evaluated values of $i_{\rm AC}$ and $R_{\rm eff}$ for the light curves of Star I and Star II in each quarter (Q2--Q16) are listed in Table \ref{tb-iac-rfg-star12}. The values of the autocorrelation index $i_{\rm AC}$, which quantify the degrees of periodicity of the light curves (see definition in Section \ref{subsec-iAC}), are also given above the corresponding $f_{\rm G}$ curves of each quarter in Figures \ref{fig04}b and \ref{fig05}b for reference. It can be seen from Figures \ref{fig04}b and \ref{fig05}b that the $i_{\rm AC}$ measure works well for the light curves of the two selected solar-type stars. If a light curve within a quarter shows strong periodicity with regular fluctuation, the $i_{\rm AC}$ value is large, implying stable magnetic features on the star's surface and thus regular rotational modulation to the light curve; if a light curve shows weak periodicity with distortions and irregular fluctuation, the $i_{\rm AC}$ value is small, implying rapid evolution of magnetic features (see detailed explanation in Section \ref{sec-introduction}). As noted in Section \ref{subsec-iAC}, the evaluated values of $i_{\rm AC}$ (see Table \ref{tb-iac-rfg-star12} or Figures \ref{fig04}b and \ref{fig05}b) are in the interval between 0 and 0.48.

In Figures \ref{fig06}a and \ref{fig06}b, we plot the variations of $i_{\rm AC}$ values with the change of quarters for Star I and Star II, respectively. Each square symbol in Figures \ref{fig06}a and \ref{fig06}b represents an $i_{\rm AC}$ value of a certain quarter: the vertical axis gives the value of $i_{\rm AC}$, the horizontal axis gives the cental time of the corresponding quarter, and the quarter number is marked on top of the square symbol. From Figures \ref{fig06}a and \ref{fig06}b it can be observed that with the change of quarters, the $i_{\rm AC}$ measure goes up to a certain maximum value and then down cyclically for both stars (note that we trace the long-term trend of $i_{\rm AC}$ and the short-term stochastic variations of $i_{\rm AC}$ in Figure \ref{fig06}a are omitted), which implies the alternate appearances of the epochs of regular fluctuation and irregular fluctuation in the light curves of the two stars (see Figures \ref{fig04} and \ref{fig05}; see also the unfolded light curves in Appendix \ref{sec-appendix-A} for an intuitive impression). We refer to each up-and-down period of $i_{\rm AC}$ variation as one $i_{\rm AC}$-cycle.

To highlight the cyclic variations of $i_{\rm AC}$, we indicate the quarters that have the maximum values of $i_{\rm AC}$ in each $i_{\rm AC}$-cycle by vertical dotted lines in Figures \ref{fig06}a and \ref{fig06}b. With the help of these vertical dotted lines, it can be identified that the plot of $i_{\rm AC}$ for Star I involves two $i_{\rm AC}$-cycles (maximum $i_{\rm AC}$ at Q3 and Q13 for each $i_{\rm AC}$-cycle; see Figure \ref{fig06}a; notice that the short-term stochastic variations of $i_{\rm AC}$ have been omitted), and the cycle length (inferred from the time interval between the vertical dotted lines) is 10 quarters (about 2.5 yr), whereas the $i_{\rm AC}$ plot for Star II involves three $i_{\rm AC}$-cycles (maximum $i_{\rm AC}$ at Q5, Q10, and Q15 for each $i_{\rm AC}$-cycle; see Figure \ref{fig06}b), and the cycle length is 5 quarters (about 1.3 yr). Since the $i_{\rm AC}$ measure of the light curve is one proxy of stellar magnetic activity (reflecting the stability of the magnetic features on the host star; see Section \ref{sec-introduction}), the cyclic variations of $i_{\rm AC}$ shown in Figures \ref{fig06}a and \ref{fig06}b imply the cyclic variations of magnetic activities on the two selected solar-type stars.

The plots of another quantitative measure of light curve, $R_{\rm eff}$ (quantifying the effective range of light-curve fluctuation; see definition in Section \ref{subsec-Reff}), with the change of quarters for Star I and Star II are shown in Figures \ref{fig06}c and \ref{fig06}d, respectively. Each triangle symbol in Figures \ref{fig06}c and \ref{fig06}d represents an $R_{\rm eff}$ value associated with a certain quarter. The vertical dotted lines employed in Figures \ref{fig06}a and \ref{fig06}b to indicate the cyclic variations of $i_{\rm AC}$ are also plotted in Figures \ref{fig06}c and \ref{fig06}d for reference. It is seen that, while the $R_{\rm eff}$ plot for Star I in Figure \ref{fig06}c shows no obvious long-term trend of variation, the $R_{\rm eff}$ plot for Star II in Figure \ref{fig06}d does display the analogous cyclic variation (denoted by $R_{\rm eff}$-cycle) as the $i_{\rm AC}$ plot in Figure \ref{fig06}b, and the phase of the $R_{\rm eff}$-cycle of Star II basically coincides with the $i_{\rm AC}$-cycle as indicated by the vertical dotted lines in Figures \ref{fig06}b and \ref{fig06}d (see also the light curves of Star II in Figure \ref{fig05}b for a more intuitive impression of the cyclic variations and phase coincidence of the two measures). As another proxy of magnetic activity, $R_{\rm eff}$ reflects the depth of stellar rotational modulation and is related to the magnetic features' size and spatial distribution on the surface of the host star (see description in Section \ref{sec-introduction}). The phase coincidence between the cyclic variations of $i_{\rm AC}$ and $R_{\rm eff}$ of Star II implies that they describe the same magnetic activity cycle on this solar-type star.

\subsection{Rotation Periods and Differential Rotations of the Two Selected Solar-type Stars \label{subsec-stars-rotation-period}}

The rotation periods of the two selected solar-type stars can be obtained by measuring the periods of the gradual periodic fluctuations in the observed light curves due to the rotational modulation. As shown in Section \ref{subsec-stars-activity-properties}, in some quarters the light curves of the two stars have very weak periodicity (represented by the low $i_{\rm AC}$ values), and the rotation periods of the stars can be hardly detected based on these quarters of light-curve data. For this reason, we use the highly periodic quarters that have the maximum values of $i_{\rm AC}$ in each $i_{\rm AC}$-cycle (indicated by the vertical dotted lines in Figures \ref{fig06}a and \ref{fig06}b) to derive the rotation periods of the two solar-type stars, that is, using Q3 and Q13 for Star I, and using Q5, Q10, and Q15 for Star II.

The generalized Lomb-Scargle (GLS) periodogram method \citep{2009A&A...496..577Z} was employed to detect the periods of the light curves. The periodogram algorithm relies mainly on the sine wave fitting \citep{1982ApJ...263..835S} and has been adopted by many authors for investigating rotation periods of Kepler stars \citep[e.g.][]{2013MNRAS.432.1203M, 2013A&A...557A..11R, 2013A&A...560A...4R}. Figure \ref{fig07} gives an example of the GLS periodograms for the light curves of Star I and Star II based on the quarters of data with the highest $i_{\rm AC}$ values (Q3 for Star I and Q10 for Star II). Figures \ref{fig07}a and \ref{fig07}b display the flux curves of Star I and Star II (in Q3 and Q10, respectively), in which the relative flux $f$ is plotted in gray and the gradual variation component of relative flux $f_{\rm G}$ is plotted in black. The corresponding GLS periodograms of the light curves are derived based on the $f_{\rm G}$ data and are displayed in Figures \ref{fig07}c and \ref{fig07}d, respectively. The period values that have the highest normalized power of sine wave fitting are regarded as the rotational periods of the stars, which are indicated by the vertical dotted lines in Figures \ref{fig07}c and \ref{fig07}d. The exact values, as well as the uncertainties of the detected periods, are given next to the vertical dotted lines ($14.40\pm0.03$ days for Q3 of Star I; $6.066\pm0.002$ days for Q10 of Star II). It is demonstrated in Figure \ref{fig07} that, based on the light curves with strong periodicity (indicated by the high $i_{\rm AC}$ values), the rotational period of the host star can be determined with good precision and accuracy.

As explained above, we use the light-curve data of Q3 and Q13 to derive the rotational period of Star I and use the light-curve data of Q5, Q10, and Q15 to derive the rotational period of Star II. For each specified quarter, we obtained a rotational period ($P_{\rm rot}$) value via the GLS periodogram method. All the derived $P_{\rm rot}$ values (with small differences for the same star) are listed in the first rows of Tables \ref{tb-rp-star1} and \ref{tb-rp-star2} (for Star I and Star II, respectively). The second rows of Tables \ref{tb-rp-star1} and \ref{tb-rp-star2} are angular velocity ($\Omega=2\pi/P_{\rm rot}$) values, which were computed based on the values of $P_{\rm rot}$. In the last columns of Tables \ref{tb-rp-star1} and \ref{tb-rp-star2}, we give the mean values of $P_{\rm rot}$ and $\Omega$ of all the processed quarters. The mean values of $P_{\rm rot}$ are just the values listed in Table \ref{tb-basic-info-star12} as the nominal rotation periods of the two Kepler stars, which are 14.73 and 6.041 days (for Star I and Star II), respectively.

The fact that there are small differences between the $P_{\rm rot}$ (or $\Omega$) values of different quarters for both Star I and Star II (see Tables \ref{tb-rp-star1} and \ref{tb-rp-star2}) implies the existence of differential rotations \citep{2005MNRAS.357L...1B, 2013A&A...560A...4R, 2014A&A...564A..50L} on the surface of the two stars. To quantify the extents of the differential rotations, we give the $\Delta P_{\rm rot}$ and $\Delta\Omega$ values in Tables \ref{tb-rp-star1} and \ref{tb-rp-star2} (see next-to-last columns), which are defined as the maximum differences between the values of different quarters. It is seen from Tables \ref{tb-rp-star1} and \ref{tb-rp-star2} that the $\Delta\Omega$ values (absolute shear of angular velocity) of Star I and Star II are 0.0188 and 0.0065 ${\rm rad}~{\rm d}^{-1}$, respectively. Dividing $\Delta\Omega$ by the mean values of $\Omega$, we get the values of relative shear $\Delta\Omega/\Omega$, which are about 0.044 and 0.0063 for Star I and Star II, respectively. Because we have no observations on the latitude distributions of the magnetic features that cause the rotational modulation, the $\Delta\Omega$ and $\Delta\Omega/\Omega$ values given above are the lower limits of the differential rotations on the two solar-type stars. Nevertheless, it can be deduced that the star with the faster rotation rate (i.e., Star II) tends to have a smaller shear of angular velocity, which is consistent with the statistical trend achieved by \citet{2013A&A...560A...4R} using large samples of Kepler stars.

\section{Activity Analysis for the Sun \label{sec-activity-sun}}

The disk-integrated radiation of the Sun has been monitored by a series of satellites and spacecrafts for decades \citep{2003GeoRL..30.1199W, 2004A&ARv..12..273F, 2005SoPh..230..129K, 2009SSRv..145..337D}. The solar light-curve data can be obtained from these observations, which can be employed for the magnetic activity analysis of the Sun by using the two quantitative measures, $i_{\rm AC}$ and $R_{\rm eff}$, as for the light curves of the two Kepler stars in Section \ref{sec-activity-stars}.

The solar light-curve data employed in this paper are the total solar irradiance (TSI) data observed by the SOHO spacecraft \citep{1995SoPh..162....1D} from the year 1996 to 2013 (18 yr in total). The instruments that recorded the TSI flux are from the Variability of solar IRradiance and Gravity Oscillations (VIRGO) investigation \citep{1997SoPh..175..267F} on SOHO. The TSI data of VIRGO on SOHO have been adopted by various authors to perform the comparative analyses between the Sun and the Kepler stars \citep[e.g.,][]{2013ApJ...769...37B, 2014A&A...568A..34B}.

Because the SOHO spacecraft that observed the TSI data is located at the Sun--Earth L1 Lagrangian point and orbits the Sun synchronously with the earth \citep{1995SoPh..162....1D}, it has a yearly cyclic variation of observation point along the heliocentric orbit, which may be reflected in the observed TSI data. For this reason, in the following reduction process for the TSI flux data, we will evaluate the two quantitative measures, $i_{\rm AC}$ and $R_{\rm eff}$, of the flux curve year by year.

Figure \ref{fig08}a gives an overview of the whole TSI flux curve from 1996 to 2013 (VIRGO TSI data set version: 6\_004\_1411). The cadence of the data set is 1 hr. The time system used in Figure \ref{fig08}a, as well as in the following figures, for the TSI data is $\rm{JD}-2,450,083.5$, where JD refers to Julian Date and the offset 2,450,083.5 is the value of the Julian Date at 00:00 UT on 1996 January 1. The TSI flux curves of different years are separated by vertical dotted lines in Figure \ref{fig08}a, and the year numbers are marked below the corresponding curves. As shown in Figure \ref{fig08}a, the long length-scale trend of the flux variation is well preserved in the TSI data.

For making a comparative analysis between the Sun and the two Kepler stars, we filtered out the long length-scale trend in the TSI data via a Fourier-based high-pass filter to simulate the observational property of the Kepler data. The result after the filtering is the PDC-like flux $F$ of the TSI data. The lower cutoff frequency of the high-pass filter was set as $1/40$ ${\rm day}^{-1}$, which is smaller than the differential rotation frequencies at all latitudes of the Sun \citep{1990SoPh..130..295H}, and thus the rotational modulation signal in the TSI data can be preserved.

The follow-up TSI data reduction procedure is similar to the Kepler data. First, we made the relative flux data $f$ by using equation (\ref{equ:f}) from the PDC-like flux $F$. Second, the gradual variation component of the relative flux, $f_{\rm G}$, was obtained from $f$ through a low-pass filter with the upper cutoff frequency being set as $1/4$ ${\rm day}^{-1}$. (This cutoff frequency was determined via visual inspection by considering the noise level and fluctuation property of the TSI flux data, which are different from the Kepler light-curve data processed in Section \ref{subsec-data-reduction}.) Third, based on the $f_{\rm G}$ time series data, the values of the two quantitative measures, $i_{\rm AC}$ and $R_{\rm eff}$ (proxies of magnetic activity), were evaluated by using equations (\ref{equ:iAC}) and (\ref{equ:R_eff}) for each year's TSI flux curve of the Sun.

Figure \ref{fig08}b displays the whole $f_{\rm G}$ flux curve of the TSI data obtained during the data reduction process. Compared with the original TSI flux curve in Figure \ref{fig08}a, the $f_{\rm G}$ curve in Figure \ref{fig08}b fluctuates around the zero value with both the long length-scale trend and the transient variation component being removed from the observed TSI flux data, and thus it can be comparable to the $f_{\rm G}$ curves of the two Kepler stars in Figures \ref{fig04}b and \ref{fig05}b. The evaluated values of $i_{\rm AC}$ and $R_{\rm eff}$ for the light curve of the Sun based on each year's $f_{\rm G}$ data (from 1996 to 2013, 18 yr in total) are listed in Table \ref{tb-iac-rfg-sun}. The $i_{\rm AC}$ values of the Sun are also given above the corresponding $f_{\rm G}$ curves of each year in Figure \ref{fig08}b for reference.

In Figure \ref{fig09}, we plot the variation of $i_{\rm AC}$ and $R_{\rm eff}$ values with time for the Sun. The square symbols in Figure \ref{fig09}a represent the values of $i_{\rm AC}$ for each year, the triangle symbols in Figure \ref{fig09}b represent the values of $R_{\rm eff}$, and the last two digits of the year numbers are marked on top of the corresponding square and triangle symbols in Figures \ref{fig09}a and \ref{fig09}b. It can be seen in Figure \ref{fig09} that both the $i_{\rm AC}$ and $R_{\rm eff}$ plots for the Sun present cyclic variations. As with the plots for the two Kepler stars in Figure \ref{fig06}, we indicate the years that have the maximum values of $i_{\rm AC}$ in each $i_{\rm AC}$-cycle of the Sun by vertical dotted lines in Figure \ref{fig09}a; the same vertical dotted lines are also plotted in Figure \ref{fig09}b for reference. With the help of these vertical dotted lines, it is found that the $i_{\rm AC}$ and $R_{\rm eff}$ values of the Sun vary in opposite phase; that is, the maxima of the $i_{\rm AC}$-cycle (specified by the vertical dotted lines in Figure \ref{fig09}) correspond to the minima of the $R_{\rm eff}$-cycle.

On the Sun, there exists the known 11 yr cycle of solar activity, which is defined by the long-term cyclic variation of sunspot number \citep{2010LRSP....7....1H}. The solar cycle can also be manifested by other indicators, such as the long length-scale trend of the TSI flux \citep{2004A&ARv..12..273F} as illustrated in Figure \ref{fig08}a, which is controlled by bright faculae rather than dark sunspots \citep{1988ApJ...328..347F}. The TSI flux data employed in this work (see Figure \ref{fig08}a) cover the solar cycle 23 (1996--2008) and the ongoing solar cycle 24 (2009--2013) \citep{2010LRSP....7....1H}. From Figure \ref{fig09} it can be seen that both the cyclic variations of $i_{\rm AC}$ and $R_{\rm eff}$ coincide with the known 11 yr solar cycle, but $i_{\rm AC}$ varies in the opposite phase, while $R_{\rm eff}$ varies in the same phase of the solar cycle (see also Figure \ref{fig08}b for a more intuitive impression).

In fact, the sunspot number that defines the solar cycle can also be regarded as a kind of proxy of magnetic activity  of the Sun based on the white-light imaging observation. Yet the sunspot number and the two measures of the light curve $i_{\rm AC}$ and $R_{\rm eff}$ reflect different aspects of the magnetic activity on the Sun: the sunspot number indicates the population of sunspots, the $R_{\rm eff}$ measure represents the size of magnetic features (mainly sunspot groups), while the $i_{\rm AC}$ measure is concerned with the stability (time evolution) of magnetic features (both sunspots and faculae). During solar cycle maximum, when the sunspot number is large \citep{2010LRSP....7....1H}, the value of $R_{\rm eff}$ is also large as shown in Figure \ref{fig09}b (same phase with solar cycle), implying big sunspot groups on the solar surface; in the meantime, the $i_{\rm AC}$ measure has a relatively low value as shown in Figure \ref{fig09}a (opposite phase with solar cycle), implying rapid evolution of sunspots and faculae. During solar cycle minimum, both the sunspot number and $R_{\rm eff}$ measure have very low values, whereas the value of $i_{\rm AC}$ is high (indicated by the vertical dotted lines in Figure \ref{fig09}), implying stable (relative to the rotation period of the Sun) magnetic features on the solar surface (see Section \ref{subsec-megentic-features} for further discussion about this issue).

From the $f_{\rm G}$ flux data of the Sun in the years 1996 and 2009, which have the maximum values of $i_{\rm AC}$ in each $i_{\rm AC}$-cycle (indicated by the vertical dotted lines in Figure \ref{fig09}a), the rotational period of the Sun can be derived by using the GLS periodogram method \citep{2009A&A...496..577Z} as for the two Kepler stars in Section \ref{subsec-stars-rotation-period}. Figure \ref{fig10} gives an example of the GLS periodograms corresponding to the light curve of the Sun in 2009 that possesses the highest value of $i_{\rm AC}$. All the derived rotational period ($P_{\rm rot}$) and angular velocity ($\Omega$) values of the Sun based on the light curves in 1996 and 2009 are listed in Table \ref{tb-rp-sun}. As anticipated, there is apparent differential rotation on the surface of the Sun (see $\Delta P_{\rm rot}$ and $\Delta\Omega$ values in Table \ref{tb-rp-sun}). The mean value of the derived rotational periods of the Sun is 27.33 days, which coincides with the value measured directly from the imaging observation \citep{1990SoPh..130..295H}. (Because the SOHO spacecraft that observed the TSI data orbits the Sun synchronously with the earth \citep{1995SoPh..162....1D}, the derived rotation period of the Sun is the synodic period and not the sidereal period as obtained from the Kepler data.)

\section{Discussion \label{sec-discussion}}

\subsection{Dominant Magnetic Feature That Causes the Periodic Light Curve \label{subsec-megentic-features}}

During solar cycle minima, the TSI flux curve of the Sun shows small amplitude (represented by low $R_{\rm eff}$ values) but apparent periodic fluctuations (represented by relatively high $i_{\rm AC}$ values; see Section \ref{sec-activity-sun}), implying that there exist long-lifetime magnetic features on the solar surface that cause the steady rotational modulation. The lifetimes of most sunspots are shorter than the rotation period of the Sun owing to their small spatial scale \citep{1997SoPh..176..249P}, not to mention the even smaller sunspots at solar minima. Thus, the sunspots appearing at solar minima tend to cause disturbance to the light curve rather than steady rotational modulation. The TSI flux curve during 1996 September--December (around solar minimum) demonstrates this situation, which is displayed in Figure \ref{fig11}a. The segment of TSI flux curve shown in Figure \ref{fig11}a presents steady periodic fluctuation as a whole, but is interrupted by a big dip on the second half of the flux curve. The phase of the big dip on the time axis dose not coincide with the troughs of the periodic fluctuation component of the light curve, which implies that the big dip and the periodic component of the light curve are caused by different kinds of magnetic features. We inferred that the big dip (disturbance to the light curve) was caused by a sunspot group that emerged on the observable disk of Sun, which eventually disappeared on the west limb along with the solar rotation; however, the long-lifetime magnetic feature that caused the periodic fluctuation of the light curve remains unclear.

To find out the exact magnetic features that are responsible for the big dip and the periodic fluctuation component in the solar light curve shown in Figure \ref{fig11}a, we draw the synoptic chart of the solar photospheric magnetograms for Carrington rotation (CR) 1916 in Figure \ref{fig11}b (white representing positive polarity of the photospheric magnetic field and black color representing negative polarity). The magnetic field data used to make the synoptic chart were observed by the Michelson Doppler Imager (MDI) instrument \citep{1995SoPh..162..129S} aboard SOHO \citep{1995SoPh..162....1D}. We choose CR 1916 because its time span (1996 November 11--December 9) just covers the portion of the TSI flux curve disturbed by the big dip, which is highlighted in gray in Figure \ref{fig11}a. This segment of the TSI flux curve covered by CR 1916 is also plotted on the synoptic magnetic field map in Figure \ref{fig11}b in white color according to the central meridian passage time given by the top horizontal axis of the chart (notice the opposite direction of the time axis in Figure \ref{fig11}b compared with the plot in Figure \ref{fig11}a). For ease of identifying the periodic fluctuation component of the light curve, in both Figures \ref{fig11}a and \ref{fig11}b we indicate the crest of the light curve just before the big dip by an upward-pointing white arrow and indicate the trough just after the big dip by a downward-pointing white arrow.

From Figure \ref{fig11}b, the exact magnetic features that correspond to each of the light curve's characteristics can be identified. The big dip in the light curve is associated with the active region (NOAA serial number: 7999) located at the center of the synoptic chart (noted by a downward-pointing black arrow in Figure \ref{fig11}b). The intense magnetic field in the active region caused the dark sunspots, and the dark sunspots in turn led to the sharp drop of the TSI flux, which forms the big dip. The crest of the light curve just before the big dip in the right half of the synoptic chart (see the upward-pointing white arrow in Figure \ref{fig11}b) is associated with an enhanced network region (noted by an upward-pointing black arrow). The enhanced network region has the enhanced network magnetic field as demonstrated in Figure \ref{fig11}b, which caused an area of concentrated bright faculae on the photosphere \citep{2006RPPh...69..563S}, and it is these bright faculae that raised the TSI flux and led to the crest of the light curve. No apparent magnetic features exist in the left half of the synoptic chart, so the TSI flux fell back to the base level and formed the trough of the light curve (indicated by the downward-pointing white arrow in Figure \ref{fig11}b).

Based on the above discussion, it can be concluded that the magnetic features associated with the periodic fluctuation component of the solar light curve in Figure \ref{fig11}a are the faculae of the enhanced network region on the surface of the Sun. Unlike dark sunspots, bright faculae affect the solar light curve by raising the solar irradiance flux. The larger spatial scale of the enhanced network region (see Figure \ref{fig11}b) favors a relatively longer lifetime of this magnetic feature \citep{1985AuJPh..38..961Z, 2000Natur.408..445S}, which tends to yield a steady rotational modulation and thus periodic light curve. In addition, the larger spatial scale of the facula zone (enhanced network region) lets the shape of the periodic light curve be more gentle and rounded (see Figure \ref{fig11}a), in contrast to the sharp (V-shaped) profile of the big dip caused by the smaller sunspots.

For the two solar-type stars observed with Kepler (see Section \ref{sec-activity-stars}), if we assume that the highly periodic light curves (characterized by high $i_{\rm AC}$ values) are also caused by the enhanced network regions and bright faculae, then the facula activities on the two solar-type stars are much stronger than the facula activities on the Sun during solar cycle minima. The highest $i_{\rm AC}$ values of Star I (in Q3) and Star II (in Q10) are 0.335 and 0.426, respectively, and the corresponding $R_{\rm eff}$ values are $0.649 \times 10^{-3}$ and $0.946 \times 10^{-3}$ (see Table \ref{tb-iac-rfg-star12}). As a comparison, the highest $i_{\rm AC}$ value of the Sun (in 2009) is just 0.162 and the corresponding $R_{\rm eff}$ value is $0.104 \times 10^{-3}$ (see Table \ref{tb-iac-rfg-sun}). The larger value of $i_{\rm AC}$ (measuring the degree of periodicity of the light curve) implies a more stable structure of enhanced network regions on the two solar-type stars and thus more regular rotational modulation, and the larger value of $R_{\rm eff}$ (measuring the fluctuation range of the light curve) implies a larger (brighter) area of faculae and thus deeper rotational modulation.

It is possible that there exist long-lifetime starspots with much larger spatial scale (than sunspots) on young and rapidly rotating solar-type stars \citep[e.g.,][]{1998A&A...330..685S, 2009A&ARv..17..251S}, which can also be responsible for the highly periodic light curves. But it is not likely for the two selected solar-type stars investigated in this paper. First, the dynamic ranges (upper limit value of $R_{\rm eff}$) of the two selected stars' light curves have the same order of magnitude as the dynamic range of the Sun (about $10^{-3}$; see the first criterion for the star selection in Section \ref{subsec-star-selection} and the $R_{\rm eff}$ values in Tables \ref{tb-iac-rfg-star12} and \ref{tb-iac-rfg-sun}); if the periodic light curves of the two selected solar-type stars were caused by starspots, to obtain the needed fluctuation range of the light curves, the size of the presumed starspots would be in the same scale as the size of the biggest sunspots on the Sun, which contradicts the hypothesis that the starspots have much larger spatial scale than sunspots, yet the larger spatial scale of starspots is essential for sustaining a stable periodic light curve. Second, superflares were seldom observed in the two selected stars' light curves (see the statistical result in the paper by \citet{2013ApJS..209....5S}), which does not give a positive support to the continuous presence of large starspot groups (which are assumed to produce the superflares) on the two solar-type stars \citep{2013PASJ...65...49S}.

Nevertheless, for the stars with a high dynamic range of light curves (or with high occurrence frequency of superflares), it is still very possible that the periodic light curves are caused by the large starspots \citep[e.g.,][]{2013PASJ...65..112N, 2013ApJ...771..127N}. (In fact, a good portion of Kepler superflare stars in the list by \citet{2013ApJS..209....5S} have the dynamic ranges that are one order of magnitude greater than the Sun.)

\subsection{Correlation between $i_{\rm AC}$ and $R_{\rm eff}$ \label{subsec-i-R-correlation}}

The activity analyses in Section \ref{sec-activity-stars} (for Star I and Star II) and Section \ref{sec-activity-sun} (for the Sun) demonstrate that the time variations of the two measures of light curves ($i_{\rm AC}$ and $R_{\rm eff}$, as proxies of magnetic activity) present different phase correlations for the two selected solar-type stars and the Sun, which represent the different magnetic activity behaviors on these stars. To describe the correlations of the two measures quantitatively, we calculated the correlation coefficients between the $i_{\rm AC}$ and $R_{\rm eff}$ values for Star I, Star II, and the Sun, respectively; the derived values are listed at the end of Tables \ref{tb-iac-rfg-star12} and \ref{tb-iac-rfg-sun}.

For the Sun, the values of $i_{\rm AC}$ and $R_{\rm eff}$ show negative correlation (correlation coefficient: $-0.29$; see Table \ref{tb-iac-rfg-sun}), which means that the two measures vary in opposite phase (see plots in Figure \ref{fig09} and description in Section \ref{sec-activity-sun}). For Star II, the two measures show positive correlation (correlation coefficient: 0.66; see Table \ref{tb-iac-rfg-star12}), which means that the two measures vary in the same phase (see plots in Figures \ref{fig06}b and \ref{fig06}d and description in Section \ref{subsec-stars-activity-properties}). For Star I, the two measures show weakest correlation (correlation coefficient: $0.10$; see Table \ref{tb-iac-rfg-star12}); while $i_{\rm AC}$ exhibits cyclic variation, $R_{\rm eff}$ varies randomly (see plots in Figures \ref{fig06}a and \ref{fig06}c).

In the process of Kepler star selection, we selected the two solar-type star samples (Star I and Star II) to intentionally let them have distinctly different rotation periods (or rotation rates; see the third criterion for the star selection in Section \ref{subsec-star-selection} and the $P_{\rm rot}$ and $\Omega$ values in Tables \ref{tb-rp-star1} and \ref{tb-rp-star2}), as we expect that stars with distinct rotation rates will have different behaviors of magnetic activity \citep{1984ApJ...279..763N, 2007LRSP....4....3G, 2010ARA&A..48..581S}. If it is assumed that the stellar activity behaviors are related to the rotation rates of stars, based on the $i_{\rm AC}$ and $R_{\rm eff}$ correlations discussed above, we can say that $i_{\rm AC}$ and $R_{\rm eff}$ show positive correlation for the faster-rotating star (Star II) and negative correlation for the slower-rotating star (the Sun), and Star I (which has the medium rotation rate and the weakest $i_{\rm AC}$ and $R_{\rm eff}$ correlation) represents a shift between the two behaviors of stellar activity. This variation of $i_{\rm AC}$ and $R_{\rm eff}$ correlation coefficients with the rotation rates (expressed by angular velocity $\Omega$) of the Sun, Star I, and Star II is plotted in Figure \ref{fig12}. (The $\Omega$ value of the Sun adopted in Figure 12 is the sidereal value, i.e., $2\pi/25.38\approx0.2476$ ${\rm rad}~{\rm d}^{-1}$, instead of the synodic value obtained in Section 4, for being compatible with the sidereal values of the two Kepler stars.)

\subsection{Stellar Activity Cycle \label{subsec-activity-cycles}}

The magnetic activity cycles on stars other than the Sun have been discovered and confirmed through various approaches, such as monitoring long-term variations of stellar photometric and chromospheric emissions \citep{1995ApJ...438..269B, 1998ApJS..118..239R, 2007ApJS..171..260L, 2008LRSP....5....2H}, tracking long-term variation of stellar X-ray luminosity \citep{2004A&ARv..12...71G, 2013A&A...553L...6S}, asteroseismology method \citep{2010Sci...329.1032G}, and Zeeman--Doppler imaging technique \citep{2011AN....332..866M}.

In this work, we rely on the signals of rotational modulation in stellar light curves to study the long-term activity variations of solar-type stars. Two quantitative measures of the modulated light curve, $i_{\rm AC}$ and $R_{\rm eff}$, are employed as the proxies of magnetic activity. The results in Sections \ref{sec-activity-stars} and \ref{sec-activity-sun} demonstrate that both the measures are able to reveal magnetic activity cycles on solar-type stars; they can be complementary or give a mutual proof result.

In the case of Star II, the time span of Kepler data covers three activity cycles (two full cycles and one half cycle; see Figures \ref{fig06}b and \ref{fig06}d). It can be seen in Figures \ref{fig06}b and \ref{fig06}d that the profiles of $i_{\rm AC}$ or $R_{\rm eff}$ variation in different cycles are not the same (see also Figure \ref{fig05}b for a more intuitive impression); each cycle has its own characteristics. Some cycles seem stronger, and some cycles are relatively weaker, just like what we had learned from the behaviors of solar cycles \citep{2010LRSP....7....1H, 2010LRSP....7....6P, 2012SoPh..281..507P}.

\section{Summary and Conclusion \label{sec-summary-conclusion}}

In this paper, we introduce an autocorrelation index, $i_{\rm AC}$, for stellar light curves as a quantitative proxy of magnetic activity for solar-type stars observed with Kepler. The concept of $i_{\rm AC}$ relies on the phenomenon of rotational modulation of stellar light curves, which is caused by the crossing of magnetic features (starspots and faculae) on a star's disk along with stellar rotation; and the value of $i_{\rm AC}$ is intended (see definition equation (\ref{equ:iAC}) in Section \ref{subsec-iAC}) to measure the degree of periodicity of the modulated light curve, which reflects the regularity of the rotational modulation and is related to the stability (and hence time evolution information) of magnetic features.

Besides $i_{\rm AC}$, we also employed $R_{\rm eff}$, the effective range of light-curve fluctuation (see definition equation (\ref{equ:R_eff}) in Section \ref{subsec-Reff}), as another proxy of stellar magnetic activity, which reflects the depth of the rotational modulation and is related to the magnetic features' size and spatial distribution. The two quantitative measures of light curves, $i_{\rm AC}$ and $R_{\rm eff}$, depict different aspects of magnetic activities on solar-type stars; they are complementary and can give a mutual proof result.

By using the two measures $i_{\rm AC}$ and $R_{\rm eff}$, we analyzed the magnetic activity properties of two carefully selected solar-type stars observed with Kepler (Kepler ID: 9766237 and 10864581, referred to as Star I and Star II, respectively). The $i_{\rm AC}$ and $R_{\rm eff}$ values were evaluated for each quarter of the light-curve data to exhibit the different activity characteristics of different quarters. The variations of the $i_{\rm AC}$ and $R_{\rm eff}$ values with the change of quarters reveal the cyclic magnetic activities on the two solar-type stars (referred to as $i_{\rm AC}$-cycle and $R_{\rm eff}$-cycle, respectively; see Figure \ref{fig06}). Based on the $i_{\rm AC}$-cycles, it is deduced that the cycle length of Star I is about 2.5 yr and the cycle length of Star II is about 1.3 yr. The rotation periods and differential rotation information of the two selected solar-type stars were derived from the quarters of light-curve data with the maximum $i_{\rm AC}$ values in each $i_{\rm AC}$-cycle (i.e., using the highly periodic light curves), and the obtained rotation periods for Star I and Star II are 14.73 and 6.041 days, respectively (more detailed results about rotation periods and differential rotations for the two Kepler stars can be found in Tables \ref{tb-rp-star1} and \ref{tb-rp-star2}).

We also applied the two measures $i_{\rm AC}$ and $R_{\rm eff}$ to the light-curve (TSI flux) data of the Sun, our nearest solar-type star. The result shows that both the time variations of $i_{\rm AC}$ and $R_{\rm eff}$ can exhibit the known 11 yr solar cycle (note that we only use the signals of rotational modulation and do not incorporate the long length-scale trend information of the absolute TSI value), which further proves the ability of $i_{\rm AC}$ and $R_{\rm eff}$ as the proxies of stellar magnetic activity.

From the activity analysis results for Star I, Star II, and the Sun, it is found that the values of $i_{\rm AC}$ and $R_{\rm eff}$ present different correlations for different stars. For Star II (with faster rotation rate), the two measures have positive correlation and vary in the same phase; for the Sun (with slower rotation rate), the two measures have negative correlation and vary in the opposite phase; for Star I (with medium rotation rate), the two measures show weakest correlation. We propose that the different $i_{\rm AC}$ and $R_{\rm eff}$ correlations represent different behaviors of stellar activity; with the increase of rotation rate, there is a shift from the negative correlation behavior to the positive correlation behavior, and Star I (owing to its medium rotation rate) is just near the transition point.

Moreover, by comparing the light curve of the Sun during solar minimum with the simultaneous synoptic chart of the solar photospheric magnetograms (see Figure \ref{fig11}), it is identified that the magnetic features that cause the steady periodic fluctuations (corresponding to higher $i_{\rm AC}$ values) of the solar light curve are the faculae of the enhanced network region on the surface of the Sun, which have much larger spatial scale than sunspots. We propose that on the two selected solar-type stars investigated in this paper, the enhanced network regions (manifested as bright facula zones) can also be candidates for the magnetic features that cause the steady rotational modulation and hence highly periodic light curves.

In the future studies, we will try to reduce the time window of the $i_{\rm AC}$ algorithm, from one quarter (currently used) to a few of rotation periods, and thus promote the time resolution of the $i_{\rm AC}$ technique. Based on this improvement, we expect to study the relations between the magnetic activity properties and the flare activity properties \citep{2014ApJ...792...67C, 2014MNRAS.445.2268P, 2015MNRAS.447.2714B, 2015ApJ...798...92W} on the solar-type stars observed with Kepler.



\acknowledgments

This paper includes data collected by the Kepler mission. Funding for the Kepler mission is provided by the NASA Science Mission directorate. The Kepler data presented in this paper (Kepler Data Release 24) were obtained from the Mikulski Archive for Space Telescopes (MAST). We acknowledge receipt of the updated data set (Version: 6\_004\_1411) of the VIRGO Experiment on SOHO from the VIRGO Team through PMOD/WRC. SOHO is a project of international cooperation between ESA and NASA. This work is jointly supported by Strategic Priority Research Program on Space Science, Chinese Academy of Sciences (grant XDA04060801), National Natural Science Foundation of China (NSFC) through grants 40890160 and 40890161, 11303051, 11403044, 11273031, and 11221063, National Basic Research Program of China (973 Program) through grant 2011CB811406, and China Meteorological Administration (grant GYHY201106011).



\appendix

\section{Unfolded Light Curves of Each Quarter for Star I and Star II \label{sec-appendix-A}}

\begin{figure}
  \epsscale{0.5}
  \plotone{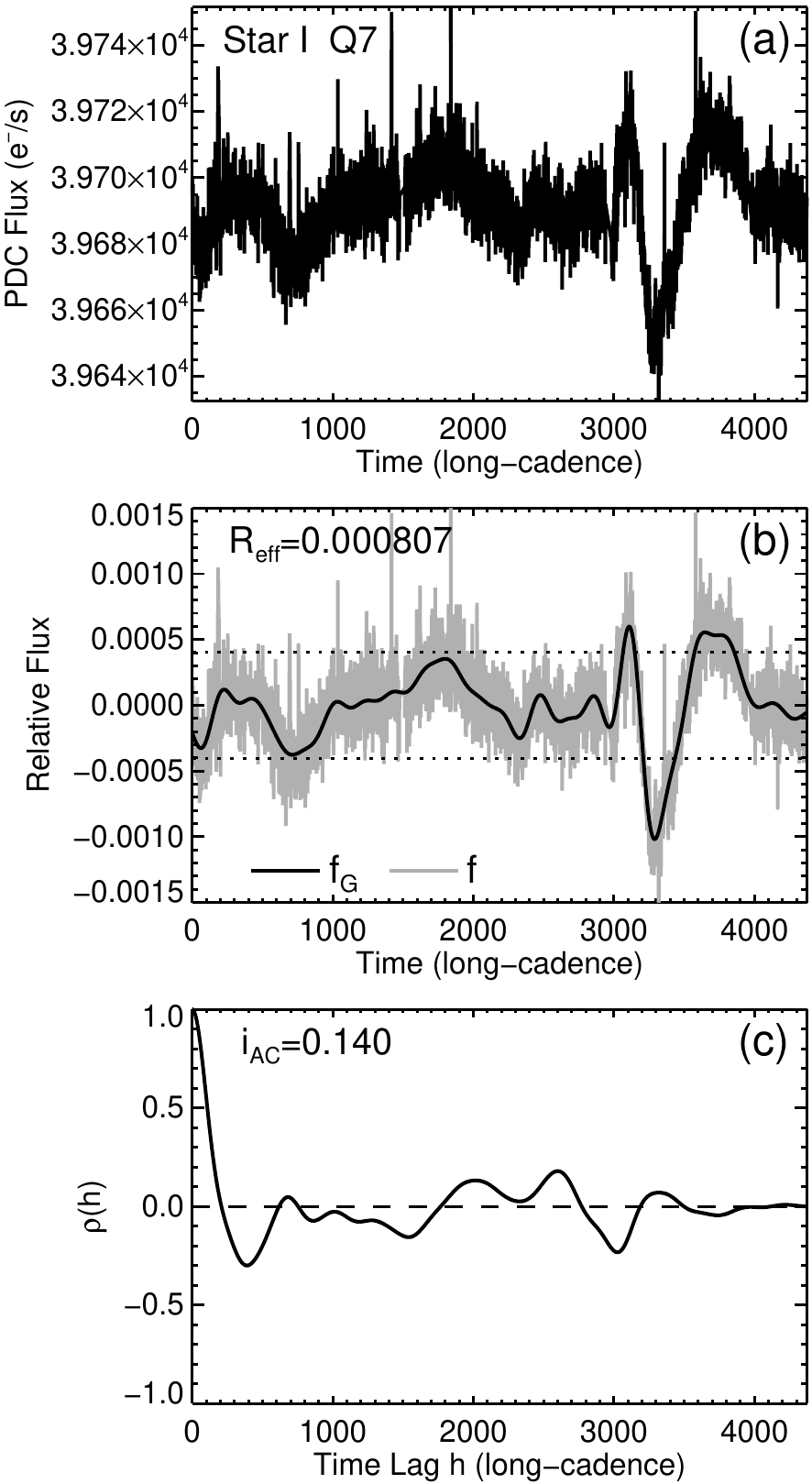}
  \caption{Illustration of data reduction procedure for one quarter of PDC flux data of Kepler. The example data are taken from Q7 of Star I. (a) Curve of the original PDC flux data. The time axis is counted from the beginning of the quarter, and the time unit is long-cadence \citep[1 long-cadence $\approx$ 29.4 minutes;][]{2010ApJ...713L.120J}. (b) Curves of the relative flux $f$ (gray) and the gradual variation component of the relative flux $f_{\rm G}$ (black). (c) Plot of the ACF, $\rho(h)$, for the $f_{\rm G}$ curve. The evaluated values of effective fluctuation range $R_{\rm eff}$ and autocorrelation index $i_{\rm AC}$ based on the $f_{\rm G}$ data (see Section \ref{subsec-data-reduction} for details) are given in panels (b) and (c), respectively. The two horizontal dotted lines in panel (b) indicate the positions of $R_{\rm eff}/2$ and $-R_{\rm eff}/2$.
  \label{fig03}}
\end{figure}

\begin{figure}
  \epsscale{1.0}
  \plotone{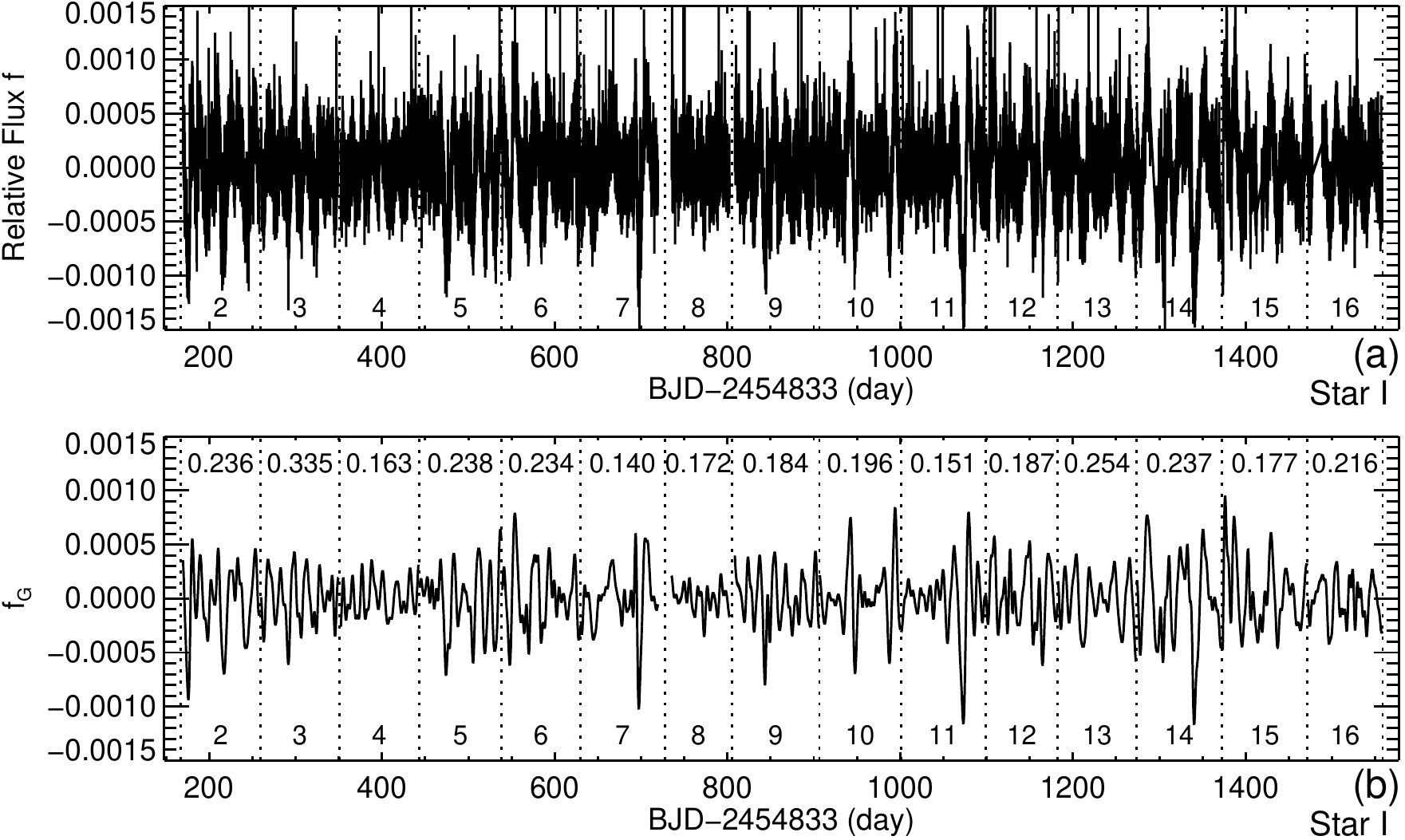}
  \caption{Overview of the whole curves of the relative flux $f$ (upper panel) and the gradual variation component of the relative flux $f_{\rm G}$ (lower panel) within Q2--Q16 for Star I. The flux curves of different quarters are separated by vertical dotted lines, and the quarter numbers are marked below the corresponding curves. The evaluated values of $i_{\rm AC}$ of each quarter are given above the corresponding $f_{\rm G}$ curves for reference.
  \label{fig04}}
\end{figure}

\begin{figure}
  \epsscale{1.0}
  \plotone{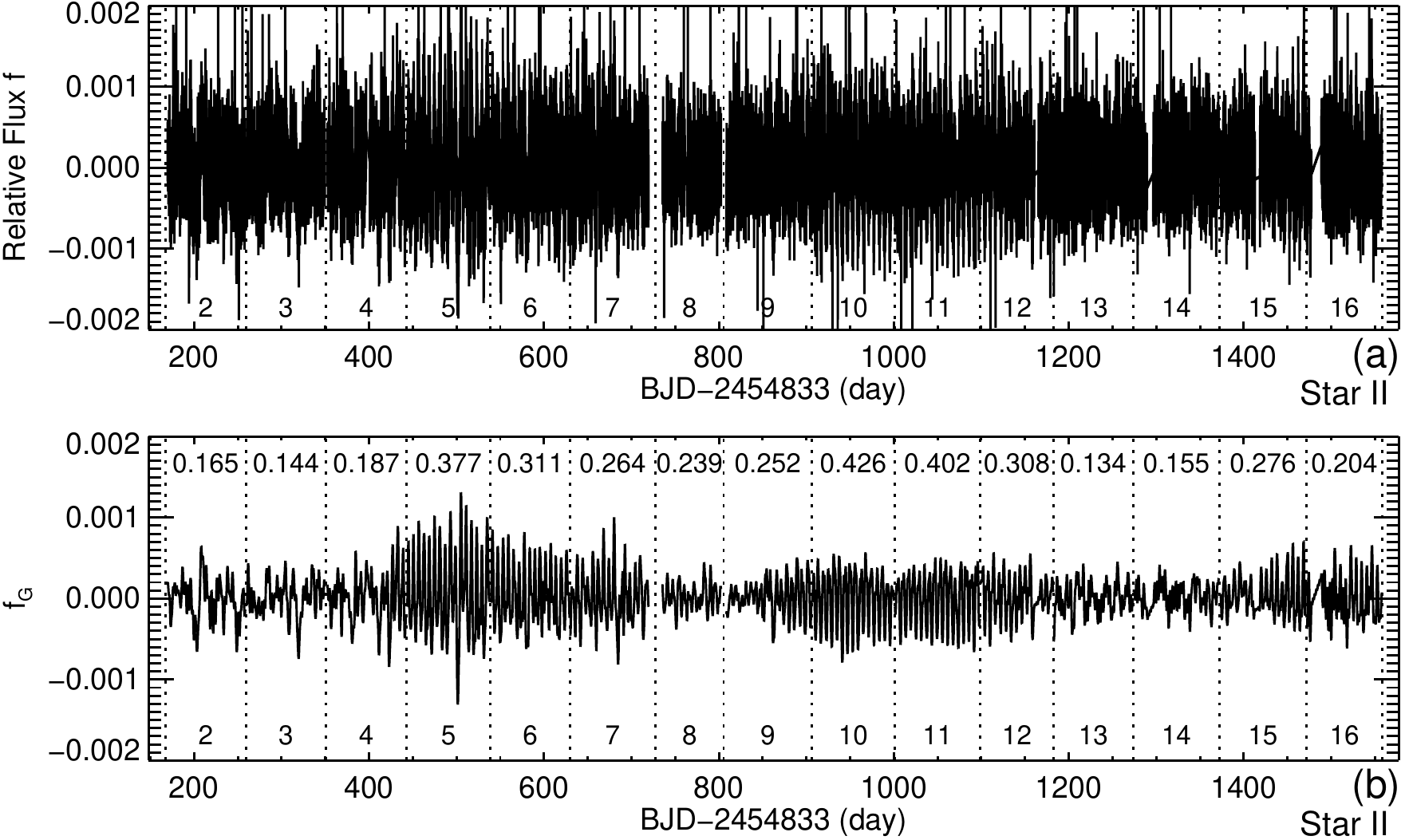}
  \caption{Same as Figure \ref{fig04}, but for Star II.
  \label{fig05}}
\end{figure}

\begin{figure}
  \epsscale{0.9}
  \plotone{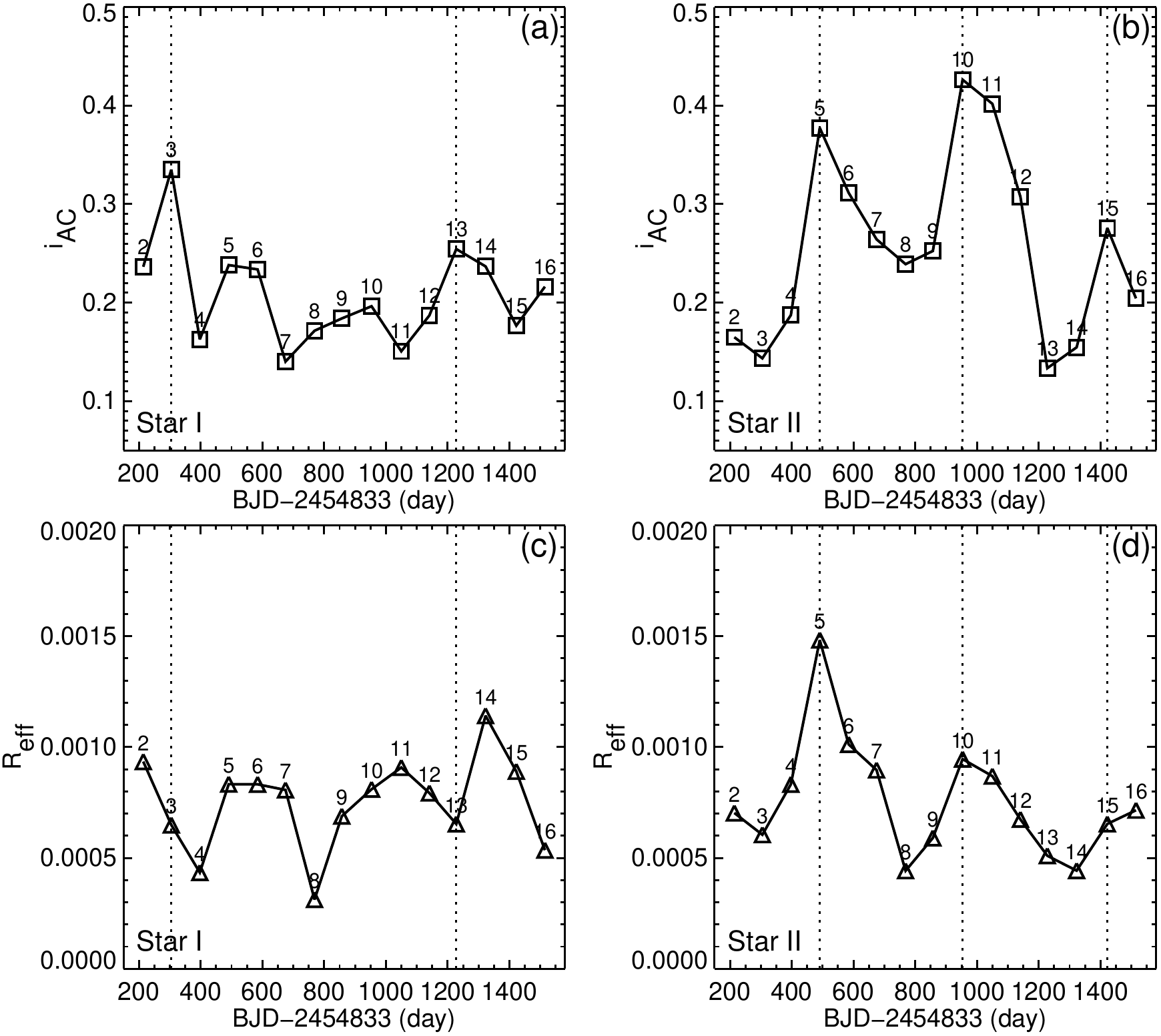}
  \caption{Plot illustrating the variations of $i_{\rm AC}$ (upper row) and $R_{\rm eff}$ (lower row) values with the change of quarters for Star I (left column) and Star II (right column). In each subplot, the vertical axis gives the values of $i_{\rm AC}$ (square symbols) or $R_{\rm eff}$ (triangle symbols) of each quarter, the horizontal coordinate gives the cental times of the quarters, and the quarter numbers are marked on top of the corresponding square and triangle symbols. The vertical dotted lines in the plots indicate the quarters that have the maximum values of $i_{\rm AC}$ in each $i_{\rm AC}$-cycle of the two stars (see Section \ref{subsec-stars-activity-properties} for details).
  \label{fig06}}
\end{figure}

\begin{figure}
  \epsscale{1.0}
  \plotone{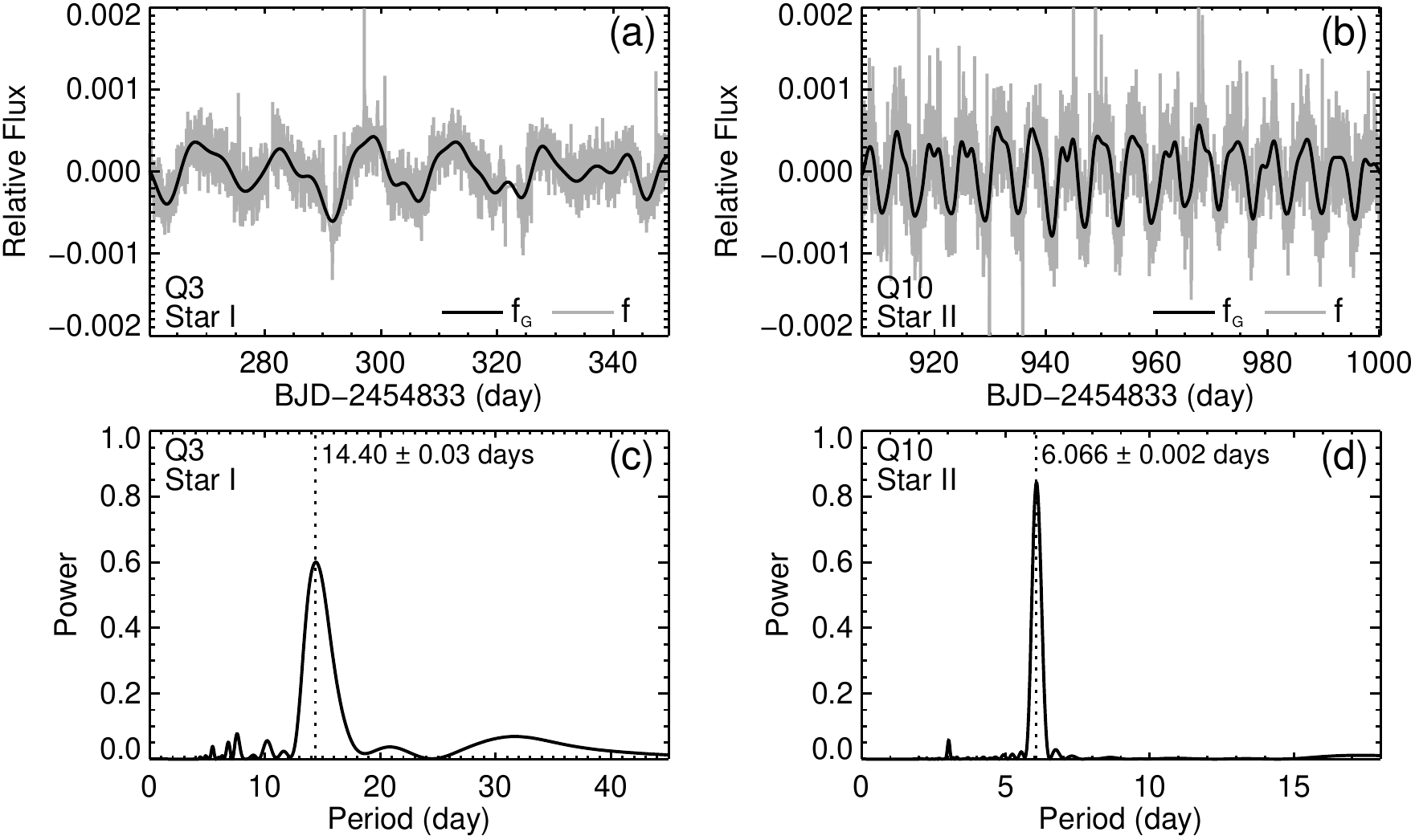}
  \caption{Flux curves of Star I and Star II in quarters that possess the highest values of $i_{\rm AC}$ (Q3 for Star I and Q10 for Star II) and their GLS periodograms for detecting the stellar rotational periods. (a) Flux curves of Star I in Q3. The relative flux $f$ is plotted in gray, and the $f_{\rm G}$ curve (gradual variation component) is in black. (b) Flux curves of Star II in Q10. (c) GLS periodogram corresponding to the $f_{\rm G}$ curve of Star I in Q3. (d) GLS periodogram corresponding to the $f_{\rm G}$ curve of Star II in Q10. The detected rotational periods are indicated by vertical dotted lines in the two periodograms; the exact values of the periods (with uncertainties) are given next to the dotted lines.
  \label{fig07}}
\end{figure}

\begin{figure}
  \epsscale{1.0}
  \plotone{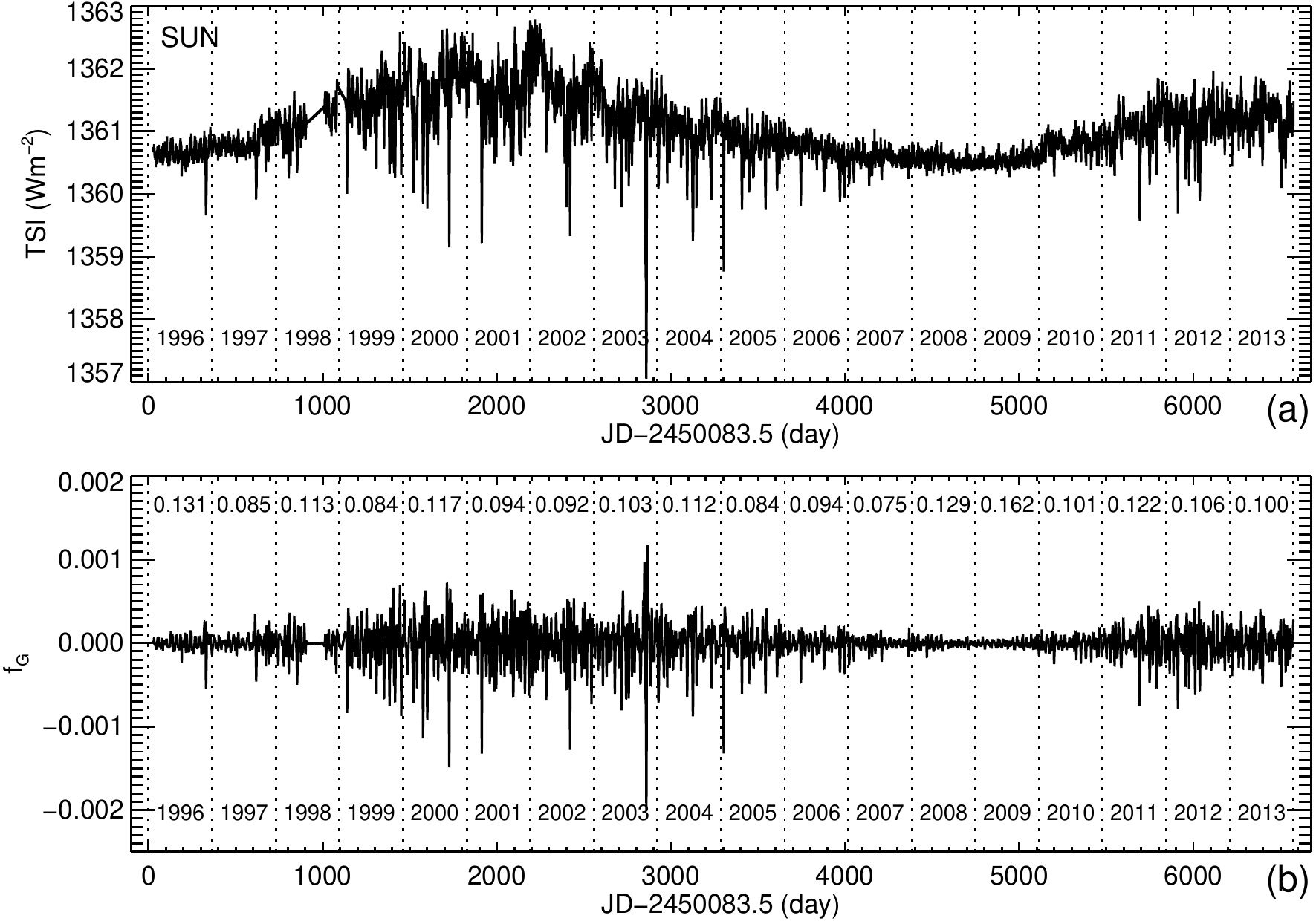}
  \caption{(a) Overview of the whole TSI flux curve from 1996 to 2013 observed with VIRGO \citep{1997SoPh..175..267F} on SOHO \citep{1995SoPh..162....1D}. The cadence of the data set is 1 hr. (b) $f_{\rm G}$ flux curve of the TSI data after the data reduction process (see Section \ref{sec-activity-sun} for details). JD of the time axis refers to Julian Date, and the offset 2,450,083.5 is the value of the Julian Date at 00:00 UT on 1996 January 1. The flux curves of different years are separated by vertical dotted lines, and the year numbers are marked below the corresponding curves. The evaluated values of $i_{\rm AC}$ of each year are given above the corresponding $f_{\rm G}$ curves for reference.
  \label{fig08}}
\end{figure}

\begin{figure}
  \epsscale{0.5}
  \plotone{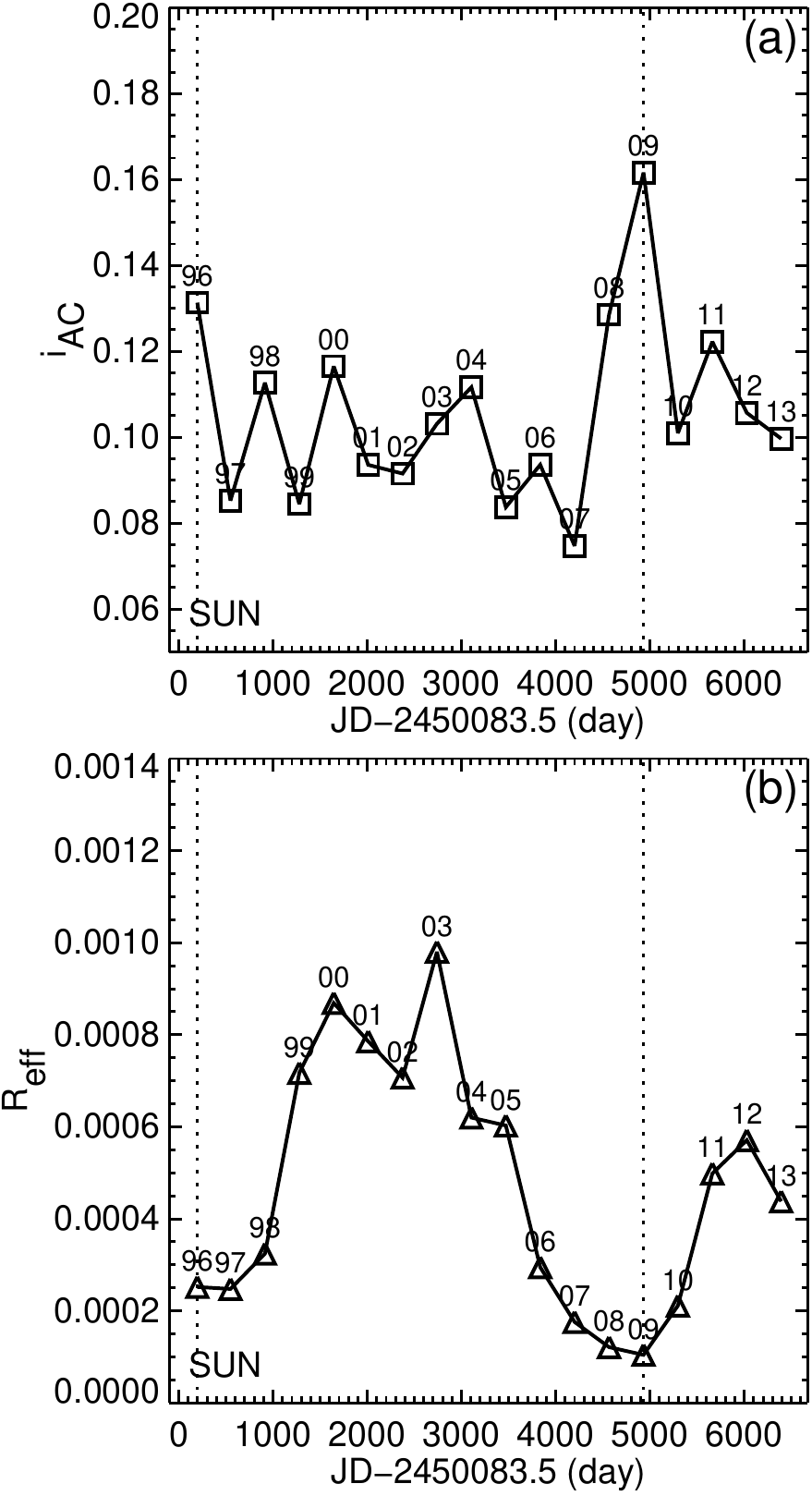}
  \caption{Plot illustrating the variation of (a) $i_{\rm AC}$ and (b) $R_{\rm eff}$ values with time for the Sun. In each subplot, the vertical axis gives the values of $i_{\rm AC}$ (square symbols) or $R_{\rm eff}$ (triangle symbols) for each year, the horizontal coordinate gives the cental times of the TSI fluxes for each year, and the last two digits of the year numbers are marked on top of the corresponding square and triangle symbols. The vertical dotted lines in the plots indicate the years that have the maximum values of $i_{\rm AC}$ in each $i_{\rm AC}$-cycle of the Sun (see Section \ref{sec-activity-sun} for details).
  \label{fig09}}
\end{figure}

\begin{figure}
  \epsscale{0.55}
  \plotone{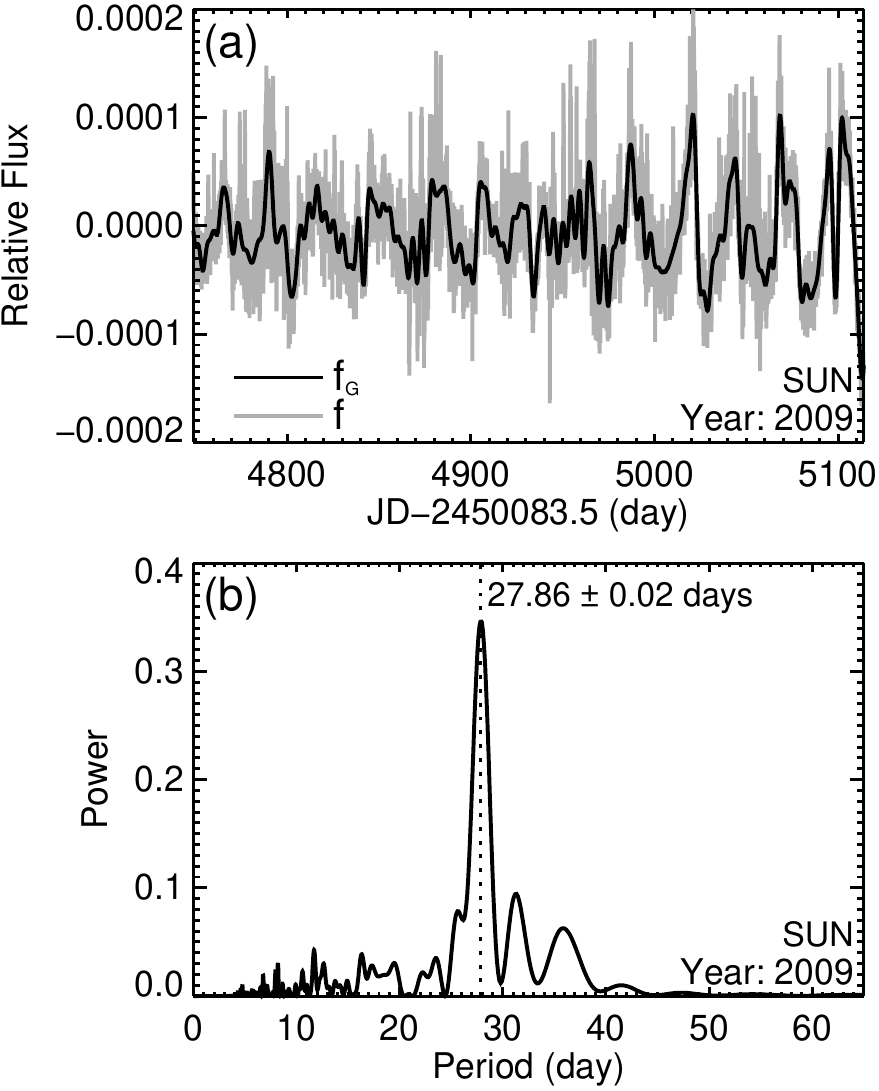}
  \caption{(a) Flux curves ($f$ and $f_{\rm G}$ curves) of the Sun in 2009. (b) GLS periodogram of the flux curve (based on $f_{\rm G}$ data) for detecting the solar rotational period. The derived rotational period is indicated by a vertical dotted line in the periodogram; the exact value of the period (with uncertainty) is given next to the dotted line.
  \label{fig10}}
\end{figure}

\begin{figure}
  \epsscale{0.8}
  \plotone{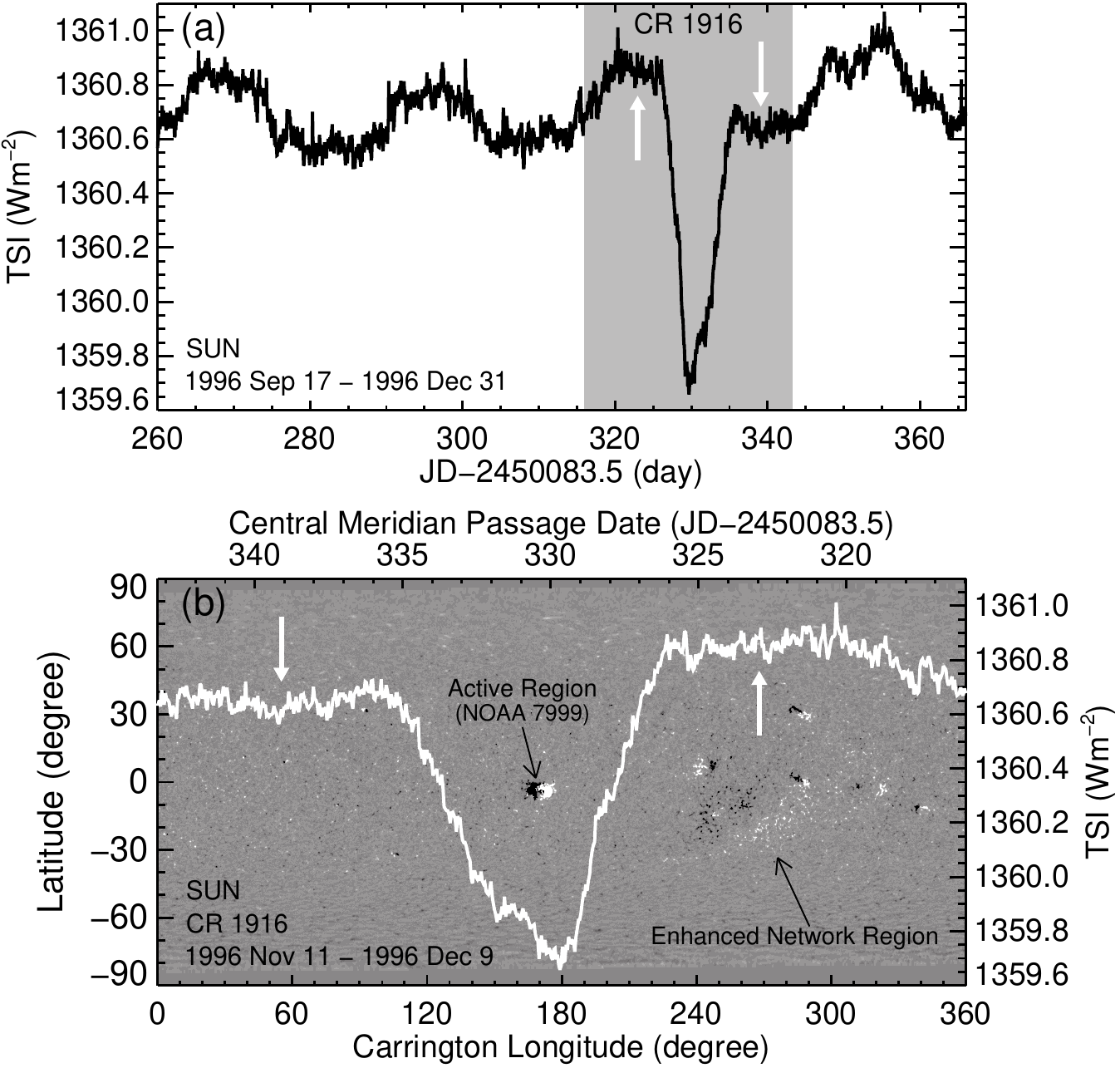}
  \caption{Comparison between the TSI flux curve of the Sun during solar minimum and the simultaneous imaging observation of solar photospheric magnetic field. (a) TSI flux curve of the Sun during solar minimum around 1996 September--December. The time period highlighted in gray corresponds to the time span of Carrington rotation (CR) 1916 (1996 November 11--December 9), which just covers the portion of the TSI flux curve disturbed by a big dip. The crest of the flux curve just before the big dip is indicated by an upward-pointing white arrow, and the trough just after the big dip is indicated by a downward-pointing white arrow. (b) Synoptic chart of the solar photospheric magnetograms for CR 1916. White color represents positive polarity, and black color represents negative polarity. The segment of TSI flux curve covered by CR 1916 (highlighted in gray in Figure \ref{fig11}a) is also plotted on the synoptic magnetic field map in white color according to the central meridian passage time given by the top horizontal axis of the chart (notice the opposite direction of this time axis compared with the plot in Figure \ref{fig11}a). The two white arrows in Figure \ref{fig11}b indicate the same crest and trough of the flux curve as in Figure \ref{fig11}a, and the two black arrows indicate the magnetic features that dominate the shape of the flux curve (see Section \ref{subsec-megentic-features} for details).
  \label{fig11}}
\end{figure}

\begin{figure}
  \epsscale{0.5}
  \plotone{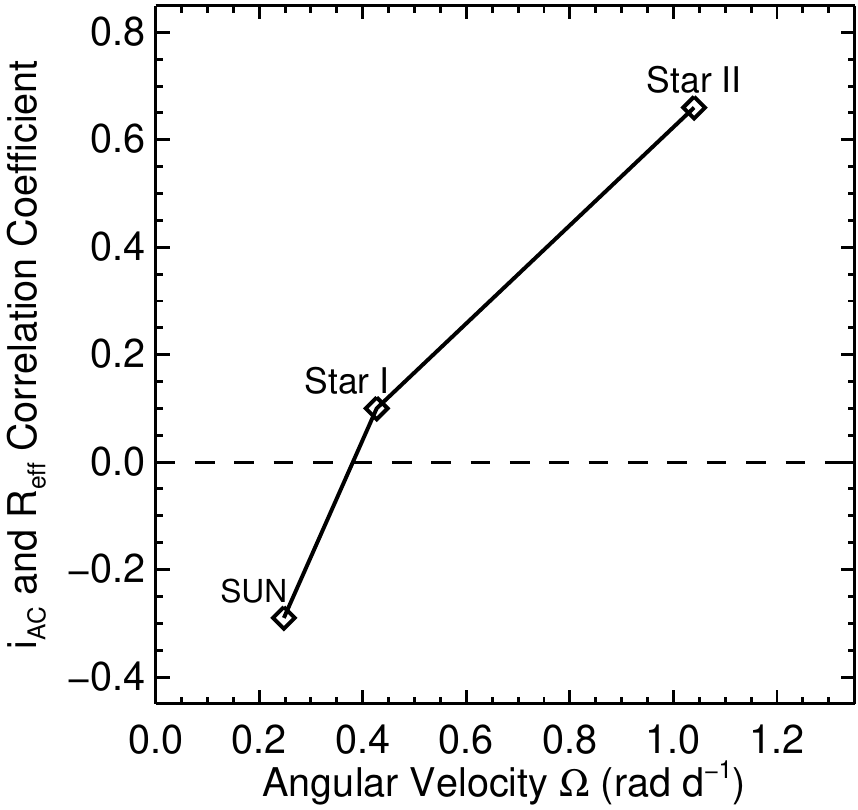}
  \caption{Plot illustrating the variation of $i_{\rm AC}$ and $R_{\rm eff}$ correlation coefficients with the rotation rates of the Sun, Star I, and Star II. The rotation rate is expressed by the angular velocity $\Omega$ in the horizontal axis. (Note that the $\Omega$ value of the Sun adopted in this plot is the sidereal value, i.e., $2\pi/25.38\approx0.2476$ ${\rm rad}~{\rm d}^{-1}$, instead of the synodic value obtained in Section \ref{sec-activity-sun} for being compatible with the sidereal values of the two Kepler stars.)
  \label{fig12}}
\end{figure}

\begin{figure}
  \epsscale{1.0}
  \plotone{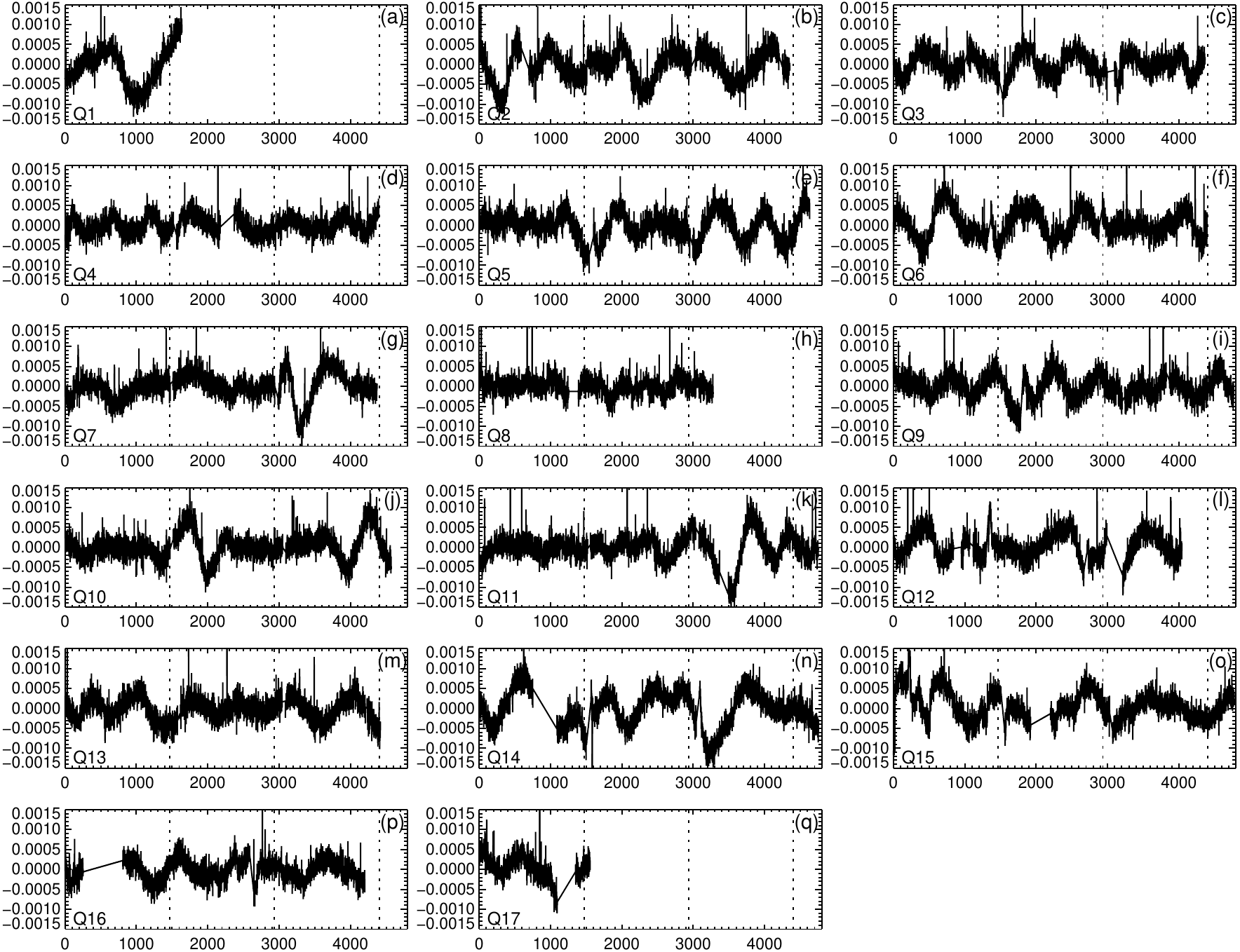}
  \caption{Unfolded light curves of each quarter for Star I. The light curves are plotted based on the relative flux data (see description in Section \ref{subsec-data-reduction}). Each subplot shows one quarter of the light curve (from Q1 to Q17); the vertical axis gives the values of relative flux $f$ (see equation (\ref{equ:f})), and the horizontal axis (time axis) gives the long-cadence number of data points, which is counted from the beginning of the quarter (the same time axis as in Figures \ref{fig03}a and \ref{fig03}b). The three vertical dotted lines in each subplot indicate the timing of 30 days, 60 days, and 90 days, respectively.
  \label{fig13}}
\end{figure}

\begin{figure}
  \epsscale{1.0}
  \plotone{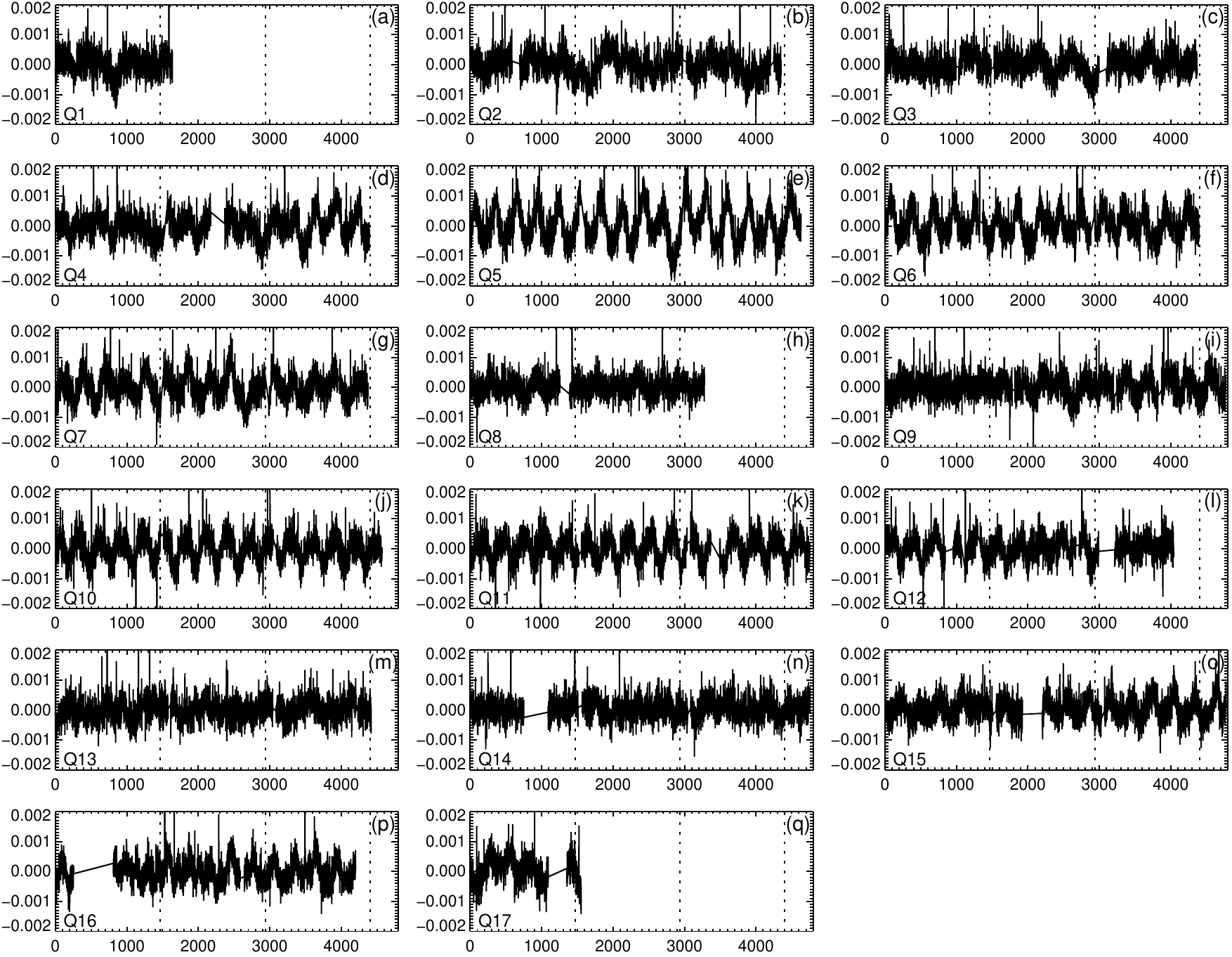}
  \caption{Same as Figure \ref{fig13}, but for Star II.
  \label{fig14}}
\end{figure}

In Figures \ref{fig04}a and \ref{fig05}a, we gave an overview of the whole light curves (in relative flux expression) for Star I and Star II, but the two plots are too compact to show the details of the light curves. For this reason, in Figures \ref{fig13} and \ref{fig14} we display the unfolded light curves of each quarter for Star I and Star II, respectively. Each subplot in Figures \ref{fig13} (for Star I) and \ref{fig14} (for Star II) shows one quarter of the light curve (from Q1 to Q17); the vertical axis gives the values of relative flux $f$ (see equation (\ref{equ:f}) in Section \ref{subsec-data-reduction}), and the horizontal axis (time axis) gives the long-cadence number of data points, which is counted from the beginning of the quarter (the same time axis as in Figures \ref{fig03}a and \ref{fig03}b). The three vertical dotted lines in each subplot of Figures \ref{fig13} and \ref{fig14} indicate the timing of 30 days, 60 days, and 90 days, respectively.





\clearpage

\clearpage

\clearpage

\clearpage

\clearpage

\clearpage

\clearpage

\clearpage

\clearpage

\clearpage

\clearpage

\clearpage

\clearpage

\clearpage

\clearpage

\begin{deluxetable}{lccccccl}
\tabletypesize{\small}
\tablecaption{Basic Information of the Two Selected Solar-type Stars Observed with Kepler
             \label{tb-basic-info-star12}}
\tablewidth{0pt}
\tablehead{
\colhead{Kepler ID} & \multicolumn{6}{c}{Stellar Properties} & \colhead{Reference Name} \\
\cline{2-7}
& \colhead{$T_{\rm eff} ({\rm K})$\tablenotemark{a}}  &  \colhead{$\log g$\tablenotemark{a}}  &
\colhead{[Fe/H]\tablenotemark{a}}  &  \colhead{$R/R_\sun$\tablenotemark{a}}  &
 \colhead{$K_p$\tablenotemark{a}} & \colhead{$P_{\rm rot} ({\rm days})$\tablenotemark{b}} & \colhead{in This Paper}
}
\startdata
 9766237  & 5674 & 4.557 & -0.147 & 0.891 & 13.913 & 14.73 & Star I  \\
 10864581 & 5426 & 4.486 & -0.694 & 0.954 & 14.909 & 6.041 & Star II  \\
\enddata
\tablenotetext{a}{The values of effective temperature (accurate to 200 K), surface gravity (accurate to 0.5 dex), metallicity (accurate to 0.5 dex.), radius, and Kepler magnitude are taken from the Kepler Input Catalog \citep[KIC;][]{2011AJ....142..112B}.}
\tablenotetext{b}{See Section \ref{subsec-stars-rotation-period} for details of the process to derive the $P_{\rm rot}$ values of the two stars.}
\end{deluxetable}

\clearpage

\begin{deluxetable}{ccccccc}
\tabletypesize{\small}
\tablecaption{Evaluated Values of $i_{\rm AC}$ and $R_{\rm eff}$ for Light Curves of Star I and Star II
              in Each Quarter \label{tb-iac-rfg-star12}}
\tablewidth{0pt}
\tablehead{
\colhead{Quarter} & &  \multicolumn{2}{c}{Star I}  &&  \multicolumn{2}{c}{Star II} \\
\cline{3-4} \cline{6-7}
 &&  \colhead{$i_{\rm AC}$}  &  \colhead{$R_{\rm eff}$}  &&  \colhead{$i_{\rm AC}$}  &  \colhead{$R_{\rm eff}$} \\
 &&                          &  $(10^{-3})$              &&                          &  $(10^{-3})$
}
\startdata
      2  &&  0.236 & 0.935  &&  0.165 & 0.703 \\
      3  &&  0.335 & 0.649  &&  0.144 & 0.605 \\
      4  &&  0.163 & 0.434  &&  0.187 & 0.832 \\
      5  &&  0.238 & 0.833  &&  0.377 & 1.483 \\
      6  &&  0.234 & 0.832  &&  0.311 & 1.012 \\
      7  &&  0.140 & 0.807  &&  0.264 & 0.897 \\
      8  &&  0.172 & 0.312  &&  0.239 & 0.442 \\
      9  &&  0.184 & 0.688  &&  0.252 & 0.590 \\
     10  &&  0.196 & 0.810  &&  0.426 & 0.946 \\
     11  &&  0.151 & 0.908  &&  0.402 & 0.869 \\
     12  &&  0.187 & 0.793  &&  0.308 & 0.674 \\
     13  &&  0.254 & 0.653  &&  0.134 & 0.510 \\
     14  &&  0.237 & 1.142  &&  0.155 & 0.443 \\
     15  &&  0.177 & 0.891  &&  0.276 & 0.654 \\
     16  &&  0.216 & 0.535  &&  0.204 & 0.715 \\
      \cline{1-7}
      && \multicolumn{5}{c}{Correlation Coefficient} \\
      \cline{3-7}
      && \multicolumn{2}{c}{$0.10$}  &&  \multicolumn{2}{c}{0.66}
\enddata
\end{deluxetable}

\clearpage

\begin{deluxetable}{lcccccc}
\tabletypesize{\small}
\tablecaption{Derived Rotation Period and Angular Velocity Values of Star I \label{tb-rp-star1}}
\tablewidth{0pt}
\tablehead{
\colhead{Measure}  &&  \multicolumn{2}{c}{Quarter}  &&  \colhead{$\Delta$\tablenotemark{a}}  &  \colhead{Mean} \\
\cline{3-4}
 &&  \colhead{Q3}  &  \colhead{Q13}  &&  &
}
\startdata
 $P_{\rm rot}$ (days) && 14.40 & 15.05 && 0.65 & 14.73  \\
 $\Omega$ (${\rm rad}~{\rm d}^{-1}$)  && 0.4363 & 0.4175 && 0.0188 & 0.4269
\enddata
\tablenotetext{a}{Difference between the values of Q3 and Q13.}
\end{deluxetable}

\begin{deluxetable}{lccccccc}
\tabletypesize{\small}
\tablecaption{Derived Rotation Period and Angular Velocity Values of Star II \label{tb-rp-star2}}
\tablewidth{0pt}
\tablehead{
\colhead{Measure}  &&  \multicolumn{3}{c}{Quarter}  &&
\colhead{$\Delta$\tablenotemark{a}}  &  \colhead{Mean} \\
\cline{3-5}
 &&  \colhead{Q5}  &  \colhead{Q10}  &  \colhead{Q15}  &&  &
}
\startdata
 $P_{\rm rot}$ (days) && 6.030 & 6.066 &  6.028 && 0.038 & 6.041  \\
 $\Omega$ (${\rm rad}~{\rm d}^{-1}$)  && 1.0420 & 1.0358  & 1.0423  && 0.0065  &  1.0400
\enddata
\tablenotetext{a}{Maximum difference between the values of Q5, Q10, and Q15.}
\end{deluxetable}

\clearpage

\begin{deluxetable}{ccccc}
\tabletypesize{\small}
\tablecaption{Evaluated Values of $i_{\rm AC}$ and $R_{\rm eff}$ for Light Curves of the Sun in Each Year
 \label{tb-iac-rfg-sun}}
\tablewidth{0pt}
\tablehead{
\colhead{Year} & &  \multicolumn{3}{c}{Sun}  \\
\cline{3-5}
 &&&  \colhead{$i_{\rm AC}$}  &  \colhead{$R_{\rm eff}$} \\
 &&&                          &  \colhead{$(10^{-3})$}
}
\startdata
   1996  &&&  0.131 & 0.252  \\
   1997  &&&  0.085 & 0.248  \\
   1998  &&&  0.113 & 0.324  \\
   1999  &&&  0.084 & 0.717  \\
   2000  &&&  0.117 & 0.869  \\
   2001  &&&  0.094 & 0.786  \\
   2002  &&&  0.092 & 0.707  \\
   2003  &&&  0.103 & 0.980  \\
   2004  &&&  0.112 & 0.620  \\
   2005  &&&  0.084 & 0.603  \\
   2006  &&&  0.094 & 0.295  \\
   2007  &&&  0.075 & 0.176  \\
   2008  &&&  0.129 & 0.121  \\
   2009  &&&  0.162 & 0.104  \\
   2010  &&&  0.101 & 0.211  \\
   2011  &&&  0.122 & 0.499  \\
   2012  &&&  0.106 & 0.571  \\
   2013  &&&  0.100 & 0.438  \\
      \cline{1-5}
      && \multicolumn{3}{c}{Correlation Coefficient} \\
      \cline{3-5}
      && \multicolumn{3}{c}{$-0.29$}
\enddata
\end{deluxetable}

\clearpage

\begin{deluxetable}{lcccccc}
\tabletypesize{\small}
\tablecaption{Derived Rotation Period and Angular Velocity Values of the Sun \label{tb-rp-sun}}
\tablewidth{0pt}
\tablehead{
\colhead{Measure} && \multicolumn{2}{c}{Year} && \colhead{$\Delta$\tablenotemark{a}} & \colhead{Mean} \\
\cline{3-4}
 && \colhead{1996} & \colhead{2009} &&  &
}
\startdata
 $P_{\rm rot}$ (days) &&  26.80  &  27.86  &&  1.06  &  27.33  \\
 $\Omega$ (${\rm rad}~{\rm d}^{-1}$)  &&  0.2344  &  0.2255  &&  0.0089  &  0.2300
\enddata
\tablecomments{Because the SOHO spacecraft that observed the solar light-curve (TSI flux) data is located at the Sun--Earth L1 Lagrangian point and orbits the Sun synchronously with the earth \citep{1995SoPh..162....1D}, the derived rotation period and angular velocity of the Sun are the synodic values and not the sidereal values as obtained from the Kepler data.}
\tablenotetext{a}{Difference between the values of the years 1996 and 2009.}
\end{deluxetable}

\end{document}